\documentclass[11pt]{article}
\usepackage{amsmath}
\usepackage[left=0.8in, right=0.8in, top=1in, textheight=9.1in]{geometry}

\usepackage{caption}

\newcommand\independent{\protect\mathpalette{\protect\independenT}{\perp}}

\def\independenT#1#2{\mathrel{\rlap{$#1#2$}\mkern2mu{#1#2}}} 
\def\bSig\mathbf{\Sigma}

\newcommand{\gte}{g_\theta(t-e-\ell)}
\newcommand{\gtr}{g_\theta(t-r-\ell)}

\newcommand{\dg}{\left( \!
\begin{array}{c}
0\\
\gte \\
\end{array} \!\right)}

\newcommand{\dge}{\left(\!
\begin{array}{c}
1 \\
\gte\\
\end{array}\!\right)}

\newcommand{\dgr}{\left(\!
\begin{array}{c}
0 \\
\gtr\\
\end{array}\!\right)}

\newcommand{\N}{\mathcal{N}}
\newcommand{\calI}{\mathcal{I}}
\newcommand{\R}{\mathcal{R}}
\newcommand{\calT}{\mathcal{T}}
\newcommand{\I}{\text{I}}

\newcommand{\pr}{\text{pr}}
\newcommand{\tilw}{\widetilde{w}}
\newcommand{\calE}{\mathcal{E}}
\newcommand{\calK}{\mathcal{K}}
\newcommand{\calU}{\mathcal{U}}

\newcommand{\tilN}{\widetilde{N}}
\newcommand{\tilY}{\widetilde{Y}}

\newcommand{\sumin}{\sum_{i=1}^n}
\newcommand{\hattheta}{\widehat{\theta}}

\newcommand{\tilbeta}{\widetilde{\beta}}
\newcommand{\tilgamma}{\widetilde{\gamma}}

\begin{document}

\begin{center}
\begin{Large}
\textbf{Estimating Vaccine Efficacy Over Time After a Randomized Study is Unblinded}
\end{Large}
\end{center}

\begin{center}
\textbf{Anastasios A. Tsiatis$^*$ and Marie Davidian$^{**}$}\\
Department of Statistics, North Carolina State University, Raleigh,
NC, USA\\
$^*$tsiatis@ncsu.edu\hspace{2in}$^{**}$davidian@ncsu.edu\\
\end{center}

\begin{center}
\textbf{Abstract}
\end{center}
  The COVID-19 pandemic due to the novel coronavirus SARS CoV-2 has
  inspired remarkable breakthroughs in development of vaccines against
  the virus and the launch of several phase 3 vaccine trials in Summer
  2020 to evaluate vaccine efficacy (VE).  Trials of vaccine
  candidates using mRNA delivery systems developed by Pfizer-BioNTech and
  Moderna have shown substantial VEs of 94-95\%, leading the US Food
  and Drug Administration to issue Emergency Use Authorizations and
  subsequent widespread administration of the vaccines.  As the trials
  continue, a key issue is the possibility that VE may wane over time.
  Ethical considerations dictate that all trial participants be
  unblinded and those randomized to placebo be offered vaccine,
  leading to trial protocol amendments specifying unblinding
  strategies. Crossover of placebo subjects to vaccine complicates
  inference on waning of VE.  We focus on the particular features of
  the Moderna trial and propose a statistical framework based on a
  potential outcomes formulation within which we develop methods for
  inference on whether or not VE wanes over time and estimation of VE
  at any post-vaccination time.  The framework clarifies assumptions
  made regarding individual- and population-level phenomena and
  acknowledges the possibility that subjects who are more or less
  likely to become infected may be crossed over to vaccine
  differentially over time. The principles of the framework can be
  adapted straightforwardly to other trials.

\vspace*{0.15in}


\clearpage

\section{Introduction}
\label{s:intro}

The primary objective of a vaccine trial is to estimate vaccine
efficacy (VE).  Typically, these trials are double-blind,
placebo-controlled studies in which participants are randomized to
either vaccine or placebo and followed for the primary endpoint, which
is often time to viral infection, on which inference on VE is based, where VE
is defined as a measure of reduction in infection risk for
vaccine relative to placebo, expressed as a percentage.  

Vaccine trials have become the focus of immense global interest as a
result of the COVID-19 disease pandemic due to the novel coronavirus
SARS-CoV-2.  The pandemic inspired unprecedented scientific
breakthroughs in the rapid development of vaccines against SARS-CoV-2,
culminating in the launch of several large phase 3 vaccine trials in
Summer 2020.  Trials in the US studying the vaccine candidates using
messenger RNA (mRNA) delivery systems developed by Pfizer-BioNTech and
Moderna began in July 2020 and demonstrated substantial evidence of
VEs of 94-95\% at interim analyses, leading the US Food and Drug
Administration (FDA) to issue Emergency Use Authorizations (EUAs) for
both vaccines in December 2020 and to the rollout
of vaccination programs shortly thereafter.

Implicit in the primary analysis in these trials is the assumption
that VE is constant over the study period and, with primary endpoint
time to infection, VE is represented by the 1 $-$ the ratio of the
hazard rate for vaccine to that for placebo, estimated based on a Cox
proportional hazards model. As the trials continue following the EUAs,
among the many issues to be addressed is the possibility that VE may
wane over time. Principled evaluation of the nature and extent of
waning of VE is of critical public health importance, as waning has
implications for measures to control the pandemic. Were all
participants in the trials to continue on their randomized assignments
(vaccine or placebo), evaluation of potential waning of VE would be
straightforward. However, once efficacy is established, ethical
considerations dictate the possibility of unblinding all participants
and offering the vaccine to those randomized to placebo. After
consultation with stakeholders, Pfizer and Moderna issued amendments
to their trial protocols specifying unblinding strategies and
modifications to planned analyses.

Crossover of placebo subjects to vaccine of necessity complicates
inference on waning of VE and has inspired recent research (Follmann
et al., 2020; Fintzi and Follmann, 2021; Lin, Zeng and Gilbert, 2021).
We propose a statistical framework within which we develop methods for
inference on whether or not VE wanes over time based on data where
subjects are unblinded and those on placebo may cross over to vaccine
and in which assumptions made regarding individual and population
phenomena are made transparent.  It is possible that subjects who are
more or less likely to become infected could be unblinded and cross
over to vaccine differentially over time, which could lead to biased
inferences due to confounding; accordingly, this possibility is
addressed explicitly in the framework.
The first author (AAT) has the privilege of serving on the Data and
Safety Monitoring Board for all US government-sponsored COVID-19
vaccine trials and is thus well-acquainted with the unblinding
approach for the Moderna trial.  Accordingly, the development is based
on the specifics of this trial, but the principles can be adapted
to the features of other trials.

In Section~\ref{s:trial}, we review the Moderna trial and the
resulting data.  We present a conceptual framework in which we
precisely define VE as a function of time post-vaccination in
Section~\ref{s:VE}.  In Section~\ref{s:framework}, we develop a formal
statistical framework within which we propose methodology for
estimation of VE and describe its practical implementation in
Section~\ref{s:implementation}.  Simulations demonstrating performance
are presented in Section~\ref{s:sim}.


\section{Clinical Trial Structure and Data}
\label{s:trial}

We first describe the timeline of the Moderna Coronavirus Efficacy
(COVE) trial (Baden et al., 2020) on the scale of calendar time.  The
trial opened on July 27, 2020 (time 0), and reached full accrual at
time $\calT_A$ (October 23, 2020).  On December 11, 2020, denoted
$\calT_P$, the FDA issued an EUA for the Pfizer vaccine, followed by
an EUA for the Moderna mRNA-1273 vaccine on $\calT_M = $ December 18,
2020.  Amendment 6 of the study protocol was issued on December 23,
2020 and specified the unblinding strategy (see Figure 2 of the
protocol) under which, starting on $\calT_U = $ December 24, 2020,
study participants are scheduled on a rolling basis over several
months for  Participant Decision clinic visits (PDCVs) at which they
will be unblinded. If originally randomized to vaccine, participants
continue to be followed; if randomized to placebo, participants can
receive the Moderna vaccine or refuse.  Let $\calT_C$ denote the time
at which all PDCVs have taken place.  The study will continue until
time $\calT_F$ at which all participants will have completed full
follow-up at 24 months after initial treatment assignment.
Assume that the analysis of vaccine efficacy using the methods in
Sections~\ref{ss:observed} and \ref{s:implementation} takes place at
time $\calT_C \leq L \leq \calT_F$, where all participants have achieved the
primary endpoint, requested to be unblinded, or attended the PDCV by
$L$.

Under this scheme, we characterize the data on a given participant as
follows.  Let $0 \leq E \leq \calT_A$ denote the calendar time at
which the subject entered the trial, $X$ denote baseline covariates,
and $A = 0$ $(1)$ if assigned to placebo (vaccine).  Denote observed
time to infection on the scale of calendar time as $U$, and
$\Delta = \I(U \leq L)$, where $\I(B) = 1$ if $B$ is true and 0
otherwise.  At $\calT_P$, availability of the Pfizer vaccine
commenced, at which point some subjects not yet infected requested to
be unblinded.  Denote by $R$ (calendar time) the minimum of (i) time
to such an unblinding, in which case $\calT_P \leq R < \calT_U$, and
define $\Gamma=1$; (ii) time of PDCV, so $ \calT_U \leq R < \calT_C$, and let
$\Gamma=2$; or (iii) time to infection, in which case $R=U$ and
$\Gamma=0$.  If $\Gamma \geq 1$ and $A=1$, so that the subject was
randomized to vaccine, s/he continues to be followed; if $A=0$, s/he
can choose to receive the Moderna vaccine, $\Psi= 1$ or refuse,
$\Psi=0$.  We distinguish the cases $\Gamma = 1$ and 2 to acknowledge
different unblinding dynamics before and after $\calT_U$.  Because a
very small number of participants requested unblinding before
$\calT_P$, and, although the protocol allows participants to refuse
unblinding at PDCV, all subjects are strongly encouraged to unblind,
we do not include these possibilities in the formulation.  

Table~\ref{f:notation} summarizes the timeline and observed data.  The
trial data are thus 
\begin{equation}\label{eq:data}
O_i =\{E_i, X_i, A_i, U_i, \Delta_i, R_i, \Gamma_i, \I(\Gamma_i \geq
1, A_i=0)\Psi_i  \},
\hspace*{0.1in} i=1,\ldots,n,
\end{equation}
independent and identically distributed (iid) across $i$.

\vspace*{-0.15in}

\begin{table}[t] 
\caption{Summary of notation.  All times are on the
    scale of calendar time, where time 0 is the start of the trial.}   
\label{f:notation}
\renewcommand{\arraystretch}{0.7} 
  \centering \begin{tabular}{ll} \hline \\*[-0.02in]
\multicolumn{2}{c}{\em Trial Milestones}\\*[0.05in]

$\calT_A$ & Full accrual reached, October 23, 2020 \\
$\calT_P$ & Pfizer granted EUA, December 11, 2020 \\
$\calT_M$ & Moderna granted EUA, December 18, 2020 \\
$\calT_U $ & Participant Decision clinic visits (PDCVs) commence,
             December 24,  2020 \\
$\calT_C$ &  PDCVs conclude \\
$\calT_F$  &  Follow-up concludes, trial ends \\
$\ell$ & Lag between initial vaccine dose and full efficacy, 6 weeks, 
         $\calT_P-\calT_A > \ell$  \\*[0.05in]

$L$ & Time of analysis of vaccine efficacy using the proposed methods;  $L >$ time at \\
       & which all  subjects have achieved the endpoint, requested unblinding, or attended\\
       & the PDCV, $L \leq \calT_F$ \\*[0.05in]

\multicolumn{2}{c}{\em Observed Data on a Trial Participant}\\*[0.05in]

$E$ &  Study entry time, $0 \leq E \leq \calT_A$ \\
$X$ & Baseline information \\
$A$ & Treatment assignment, placebo, $A=0$, or vaccine, $A=1$ \\
$U, \Delta$ & Time to infection, indicator of infection by time $L$, $\Delta = \I(U \leq L)$\\
$R, \Gamma$ & Time to requested unblinding, PDCV/requested unblinding, or infection, whichever \\
                    & comes first \\*[0.1in]
&\hspace{0.1in}$\Gamma=0$: $R=U$, infection occurs before requested/offered unblinding \\
&\hspace{0.1in}$\Gamma=1$: $R =$ time to requested unblinding,   $\calT_P \leq R < \calT_U$\\
&\hspace{0.1in}$\Gamma=2$: $R =$ time to PDCV or requested unblinding,
  $\calT_U  \leq R < \calT_C$ \\*[0.05in]

$\Psi$  & If $A=0$, $\Gamma \geq 1$, indicator or whether subject receives
         Moderna vaccine, $\Psi=1$, \\
        & or refuses, $\Psi=0$ \\*[0.05in] \hline
\end{tabular}
\end{table}

\section{Conceptualization of Vaccine Efficacy}
\label{s:VE}

Similar to Halloran, Longini, and Struchiner (1996) and Longini and
Halloran (1996), we consider the following framework in which to
conceptualize vaccine efficacy.  The study population, comprising
individuals for which inference on vaccine efficacy is of interest, is
that of individuals susceptible to infection, represented by the trial
participants.  There is a population of individuals outside the trial
with which trial participants interact, assumed to be much larger than
the number of participants, so that interactions among
participants are much less likely than interactions with the outside
population.  The probability that a trial participant will become
infected at calendar time $t$ depends on three factors: $c(t)$, the
contact rate, the number of contacts with the outside population per
unit time; $p(t)$, the prevalence of infections in the outside
population at $t$; and $\pi(t)$, the transmission probability at $t$,
the probability a susceptible individual in the study population will
become infected per contact with an infected individual from the
outside population.  Dependence of $\pi(t)$ on time acknowledges the
emergence of new variants of the virus, which may be be more or less
virulent, as in the COVID-19 pandemic.  Assuming random mixing,
$p(t) c(t)$ is the contact rate at time $t$ with infected individuals,
and the infection rate at time $t$ is $p(t) c(t) \pi(t)$.

We adapt this framework to the COVID-19 pandemic.  The prevalence rate
in the pandemic can vary substantially in time and space, so denote by
$S$ the trial site at which a participant is enrolled, and let
$p(t,s)$ be the prevalence at time $t$ at site $S=s$.  Although
$p(t,s)$ varies by $t$ and $s$, assume it is unaffected by the
individuals in the trial and thus represents an external force.  We
view the contact rate as individual specific; accordingly, for an
arbitrary individual in the study population, let the random variables
$\{c_0^b(t),c_1^b(t),c_0^u(t), c_{1\ell}^u(t),c_1^u(t)\}$ denote
potential contact rates.  These potential outcomes can be regarded as
individual-specific behavioral characteristics of trial participants,
where some may be more careful and make fewer contacts while others
take more risks, and behavior can vary over time and by vaccination and blinding
status.  Here, $c_a^b(t)$ is the contact rate at time $t$ if the
individual were to receive vaccine, $a=1$, or placebo, $a=0$, and be
blinded to this assignment; by virtue of blinding, it is reasonable to
take $c_1^b(t)=c_0^b(t)=c^b(t)$.  The Moderna vaccine is administered
in two doses, ideally 4 weeks apart, and is not thought to
achieve full efficacy until 2 weeks following the second dose.  Thus,
letting $\ell$ denote the lag between initial dose and full
efficacy, $c^u_{1\ell}(t)$ and $c^u_1(t)$ reflect behavior
of an individual who is unblinded and vaccinated in the periods prior
to $\ell$ and after $\ell$, respectively, allowing for unblinded
vaccinees to, e.g., behave more cautiously before full efficacy is
achieved.  The rate $c_0^u(t)$ reflects behavior of an
unblinded individual on placebo and does not play a role in the
development.  Similar to the stable unit treatment value assumption
(Rubin, 1980), assume that $c_{1\ell}^u(t)$ and $c^u_1(t)$ are the same
whether the individual was randomized to vaccine and unblinded before
$t$ or was randomized to placebo and subsequently unblinded and
crossed over to vaccine before $t$.

Finally, for an arbitrary participant, let the random variable
$\pi_0(t)$ be the potential individual-specific transmission
probability per contact at $t$ if s/he were to receive placebo, and let
$\pi_1(t,\tau)$ be the same if s/he were to receive vaccine and have
been vaccinated for $\tau \geq 0$ units of time.  As we now demonstrate, this
formulation allows us to represent VE as a function of $\tau$ and thus
consider whether or not VE wanes over time since vaccination.

With the set of potential outcomes for an arbitrary individual in the
study population who enrolls at site $S$ thus given by
$\{c^b(t),c_0^u(t),c_{1\ell}^u(t), c_1^u(t)~t>0,\pi_0(t),\pi_1(t,\tau),~\tau\geq 0\}$, the
infection rate in the study population at calendar time $t$ if all
individuals were to receive placebo and be blinded to that assignment
is $\calI_0^b(t) = E\{p(t,S)c^b(t)\pi_0(t)\}$; likewise, the infection
rate at $t$ if all individuals were to receive vaccine at time
$t-\tau$ and be blinded to that assignment is
$\calI_1^b(t,\tau) = E\{p(t,S)c^b(t)\pi_1(t,\tau)\}$.  The relative
infection rate at $t$ is then
\begin{equation}
\R^b(t, \tau) = \frac{\calI_1^b(t,\tau)}{\calI_0^b(t)}= 
\frac{E\{p(t,S)c^b(t)\pi_1(t,\tau)\}}{E\{p(t,S)c^b(t)\pi_0(t)\}}.
\label{eq:Rbt}
\end{equation}
 Accordingly, vaccine efficacy at time $t$ after vaccination at
 $t-\tau$ is $VE(t,\tau) = 1 - \R^b(t, \tau)$, reflecting the
 proportion of infections at $t$ that would be prevented if the study
 population were vaccinated and on vaccine  for $\tau$ units of time
 during the blinded phase of the study.  
 
 In the sequel, we assume that $\R^b(t, \tau)$ and thus $VE(t,\tau)$
 depend only on $\tau$ and write $\R^b(\tau)$ and
 $VE(\tau) = 1-\R^b(\tau)$.  This assumption embodies the belief that,
 although infection rates may change over time, the relative effect of
 vaccine to placebo remains approximately constant and holds if (i)
 $\{\pi_1(t,\tau),\pi_0(t)\}\independent \{S,c^b(t)\}|X$, where
 $\independent$ means ``independent of'' and this independence is
 conditional on $X$; and (ii)
 $E\{\pi_1(t,\tau)|X\}/E\{\pi_0(t)|X\}=q(\tau)$, so does not depend on
 $t$ and $X$.  Condition (i) reflects the interpretation of
 $\pi_1(t,\tau)$ and $\pi_0(t)$ as inherent biological characteristics
 of an individual, whereas $S$ and $c^b(t)$ are external and
 behavioral characteristics, respectively; thus, once common
 individual and external baseline covariates are taken into account,
 biological and geographic/behavioral characteristics are unrelated.
 Condition (ii) implies that, although new viral variants may change
 transmission probabilities under both vaccine and placebo over time,
 this change stays in constant proportion, and this proportion is
 similar for individuals with different characteristics.  Further
 discussion is given in Section~\ref{s:discuss} and Appendix B.

 Within this framework, the goal of inference on waning of VE based on
 the data from the trial can be stated precisely as inference on
 $VE(\tau) = 1-\R^b(\tau)$, $\tau \geq \ell$, so reflecting VE
 after full efficacy is achieved.   It is critical to recognize that, like
 estimands of interest in most clinical trials, $VE(\tau)$ represents
 VE at time since vaccination $\tau$ under the original conditions of
 the trial, under which all participants are blinded.  The challenge
 we address in subsequent sections is how to achieve valid inference on
 $VE(\tau)$, $\tau \geq \ell$, using data from the modified trial in which blinded
 participants are unblinded in a staggered fashion, with placebo
 subjects offered the option to receive vaccine.

 We propose a semiparametric model within which we cast this
 objective.  Let $\calI_{1\ell}^u(t,\tau) =
 E\{p(t,S)c_{1\ell}^u(t)\pi_1(t,\tau)\}$, $\tau < \ell$, and 
$\calI_1^u(t,\tau) = E\{p(t,S)c_1^u(t)\pi_1(t,\tau)\}$, $\tau \geq \ell$, be the
infection rates in the study population at $t$ if all individuals were
to receive vaccine at time $t-\tau$ and be unblinded to that fact.
 Analogous to (i) above, assume that
 $\{\pi_1(t,\tau),\pi_0(t)\}\independent \{S, c^u_{1\ell}(t),c^u_1(t)\}|X$,
 and continue to assume condition (ii).  Then, for two values
 $\tau_1,\tau_2$ of $\tau$,  it is straightforward that (see 
 Appendix A) 
\begin{equation}
\frac{\calI_{1\ell}^u(t,\tau_1)}{\calI_{1\ell}^u(t,\tau_2)} = 
\frac{\R^b(\tau_1)}{\R^b(\tau_2)}, \hspace*{0.1in} \tau_1, \tau_2 <\ell;
\hspace*{0.25in}
\frac{\calI_1^u(t,\tau_1)}{\calI_1^u(t,\tau_2)} = 
\frac{\R^b(\tau_1)}{\R^b(\tau_2)}, \hspace*{0.1in} \tau_1, \tau_2 \geq \ell.
\label{eq:tau12}
\end{equation}
Defining 
$\calI_{1\ell}^u(t) = \calI_{1\ell}^u(t,0)= E\{
p(t,S)c_{1\ell}^u(t)\pi_1(t, 0)\}$ and
$\calI_1^u(t) = \calI_1^u(t,\ell) = E\{ p(t,S)c_1^u(t)\pi_1(t,
\ell)\}$, by (\ref{eq:tau12}) with $\tau_1=\tau$ and $\tau_2=0$
($\ell$) on the left (right) hand side, the infection rates at $t$ if
all individuals in the study population were unblinded and to receive
vaccine at time $t-\tau$ are
\begin{equation}
\calI_{1\ell}^u(t,\tau) = \calI_{1\ell}^u(t)
\frac{\R^b(\tau)}{\R^b(0)}, \,\,\, \tau < \ell; \hspace*{0.25in}
\calI_1^u(t,\tau) = \calI_1^u(t) \frac{\R^b(\tau)}{\R^b(\ell)}, \,\,\,
\tau \geq \ell.
\label{eq:def2}
\end{equation}
Likewise, from (\ref{eq:Rbt}), the infection rate at $t$ if all individuals in the study
population were blinded and to receive vaccine at time $t-\tau$ is
\begin{equation}
\calI_1^b(t,\tau) = \calI_0^b(t) \R^b(\tau).
\label{eq:def1}
\end{equation}

We now represent the infection rate ratio $\R^b(\tau)$ as
\begin{equation}
\R^b(\tau; \theta) = \exp\{\zeta(\tau)\} \I(\tau < \ell) + 
\exp\{ \theta_0 + g(\tau-\ell; \theta_1)\} \I(\tau \geq \ell),
\hspace*{0.15in} \theta = (\theta_0, \theta_1^T)^T,
\label{eq:semimodel}
\end{equation}
where $\zeta(\tau)$ is an unspecified function of $\tau$; 
$\theta_0$ and $\theta_1$ are real- and vector-valued
parameters, respectively; and $g(u; \theta_1)$ is a real-valued
function of such that $g(0; \theta_1)=0$ for all $\theta_1$ and
$g(u; 0) = 0$.  For example, taking
$g(u; \theta_1) = \theta_1 u$ yields
$\R^b(\tau; \theta) = \exp\{\theta_0 +\theta_1(\tau-\ell)\}$, $\tau
\geq \ell$, in which case
$\theta_1 = 0$ implies that $VE(\tau) = 1-\R^b(\tau)$, $\tau \geq \ell$, does not change
with time since vaccination, and $\theta_1 > 0$ indicates that
$VE(\tau)$ decreases with increasing $\tau$; i.e., exhibits waning.
More complex specifications of $g(u; \theta_1)$ using splines (e.g.,
Fintzi and Follmann, 2021)  or piecewise constant functions could be made;
e.g., for $v_1< v_2 \leq L$, 
\begin{equation}
g(u; \theta_1) = \theta_{11} \I(v_1 < u \leq v_2) + \theta_{12} \I(u
> v_2), \hspace*{0.10in} \theta_1 =
(\theta_{11},\theta_{12})^T.
\label{eq:piecewise}
\end{equation}

Under this model, (\ref{eq:def1}) and
(\ref{eq:def2}) can be written as
\begin{equation}
\label{eq:def3}
\begin{aligned} 
&\,\,\,\calI^b_1(t, \tau) = \calI^b_0(t) \big[ \exp\{\zeta(\tau)\} \I(\tau <
\ell) + \exp\{ \theta_0 + g(\tau-\ell;  \theta_1)\} \I(\tau \geq\ell)\big], \\*[-0.03in]
&\calI^u_{1\ell}(t, \tau) = \calI^u_{1\ell}(t) \exp\{\zeta(\tau)\}, \,\, \tau
<\ell, \hspace*{0.15in} \calI^u_1(t, \tau) = \calI^u_1(t) \exp\{
g(\tau-\ell; \theta_1)\}, \,\, \tau \geq \ell. 
\end{aligned}
\end{equation}
Thus, to estimate $VE(\tau)$ for any $\tau$ and make inference on
potential waning of VE, we must develop a principled approach to
estimation of $\theta$ based on the data from the modified trial in
which participants are unblinded and those on placebo may cross over
to vaccine.

\vspace*{-0.2in}

\section{Statistical Framework}
\label{s:framework}

\vspace*{-0.1in}

\subsection{Motivation}
\label{ss:motivation}

Estimation of $VE(\tau)$, equivalently $\R^b(\tau)$, would be
straightforward for any $\tau\geq \ell$ over the entire follow-up
period if all participants remained on their assigned treatments
throughout the trial.  However, subjects randomized to placebo have
the option to cross over to vaccine on or after $\calT_P$.  For
$\tau <\calT_P$, it is possible to estimate $\R^b(\tau)$ because, due
to randomization, for $t < \calT_P$ we have representative samples of
blinded subjects on vaccine and placebo and thus information on
$\calI^b_1(t,\tau)$ and $\calI^b_0(t)$, so can estimate $\theta_0$ and
components of $\theta_1$ identified for such $\tau$; e.g., in
(\ref{eq:piecewise}) depending on the values of $v_1$ and $v_2$.  At
$\calT_P \leq t < \calT_C$, the data comprise a mixture of blinded and
unblinded participants, where, within the latter group, those on
placebo may have crossed over to vaccine.  Here, information, albeit
diminishing during the interval $[\calT_P,\calT_C)$, on
$\calI^b_1(t,\tau)$ and $\calI^b_0(t)$ is available from those
participants not yet unblinded, which contributes to estimation of
$\theta_0$ and components of $\theta_1$.  Information is also
available on $\calI^u_1(t,\tau)$ from individuals who were originally
randomized to vaccine and provide information on longer $\tau$, and
from individuals who recently crossed over to vaccine and provide
information on shorter $\tau$.  For $t \geq \calT_C$, there are no
longer blinded participants, so that information is available only on
$\calI^u_1(t,\tau)$.  For these latter groups, for longer $\tau_1 \geq\ell$ and
shorter $\tau_2\geq \ell$,
$\calI^u_1(t,\tau_1)/\calI^u_1(t,\tau_2) =
\exp\big[g\{\tau_1-\ell;\theta_1\}-g\{\tau_2-\ell;\theta_1\}\big]$, and, because
of the mixture of times since vaccination, $\theta_1$ can be fully
estimated.

Through the following potential outcomes formulation and under
suitable assumptions, in the next several sections we develop an
approach to estimation of $\theta$ based on the observed data
(\ref{eq:data}) that embodies the foregoing intuitive principles.


\subsection{Potential outcomes formulation}
\label{ss:potential}

Denote by $T^*_0(e,r)$ the potential time to infection on the scale of
patient time for an arbitrary individual in the study population if
s/he were to enter the trial at calendar time $e$, receive placebo and
be blinded to that fact, and, if not infected by calendar time $r$, be
unblinded and cross over to vaccine at $r$.  Let
$T^*_0(e) = T^*_0(e,\infty)$, if s/he is never crossed over to receive
vaccine.  Similarly, define $T^*_1(e,r)$ to be the potential time to
infection (patient time scale) for an arbitrary individual if s/he
were to enter the trial at $e$, receive vaccine and be blinded to that
fact, and, if not infected by $r$, be unblinded at $r$; and define
$T^*_1(e) = T^*_1(e, \infty)$.  We make the consistency assumptions
that $T^*_0(e,r) = T^*_0(e)$ if $T^*_0(e) < r$ and
$T^*_1(e,r) = T^*_1(e)$ if $T^*_1(e) < r$.  For $a=0, 1$, denote the
hazard at calendar time $t$, $t > e$, by
\begin{equation}
\lambda_a(t, e, r) = \lim_{dt \rightarrow 0} \pr\{ t \leq T^*_a(e,r) +
e < t+dt \,|\, T^*_a(e,r) +e \geq t\}, \hspace*{0.15in} a = 0, 1,  
\label{eq:hazarda}
\end{equation}
where the addition of $e$ induces a shift from patient to calendar
time.  Denote the set of all potential outcomes as 
$$W^*=\{T^*_0(e,r), T^*_1(e,r) ;e>0, r>e\}.$$

The development in Section~\ref{s:VE} is in terms of infection rates
at the individual-specific and population levels.  Population-level
hazard rates such as (\ref{eq:hazarda}) are not equivalent to
population-level infection rates.  However, we argue in Appendix C
that, because the probabilities of
infection under vaccine and placebo during the course of the trial are
small, population-level hazard rates and population-level infection
rates are approximately equivalent; this assumption is implicit in the
standard primary analysis noted in Section~\ref{s:intro}.    Thus, to reflect this, we use
familiar notation and write
$\lambda^b(t) =\calI^b_0(t)$, $\lambda^u_\ell(t) =
\calI^u_{1\ell}(t)$,  and $\lambda^u(t) = \calI^u_1(t)$.  Under
these conditions, using (\ref{eq:def3}), we can write for $t > e$ 
\begin{align}
  \lambda_0(t,e,r) &= \lambda^b(t) \I(t < r) + \lambda_\ell^u(t)
                   \exp\{ \zeta(t-r) \} \I(t-r < \ell) \nonumber \\*[-0.12in]
&\hspace*{0.5in}+ \lambda^u(t) \exp\{ g(t-r-\ell; \theta_1)\} \I(t-r
  \geq \ell), \label{eq:lam0r} 
\end{align}
\vspace*{-0.2in}
\begin{align}
\lambda_1(t,e,r) &= \lambda^b(t) \big[ \exp\{ \zeta(t-e) \} \I(t-e <\ell) 
+ \exp\{\theta_0 +g(t-e- \ell; \theta_1) \} \I(t-e \geq \ell) \big]\I(t < r) \nonumber \\*[-0.12in]
&\hspace*{0.5in}+ \lambda^u(t) \exp\{ g(t-e-\ell;   \theta_1)\} \I(t \geq r), \label{eq:lam1r}
\end{align}
where (\ref{eq:lam1r}) follows because $r \geq \calT_P$,
$e \leq \calT_A$, $\calT_P-\calT_A > \ell$. Define the counting
processes for infection by
$N^*_a(t,e,r) = \I\{ T^*_a(e,r) + e \leq t\}$ and
$N^*_a(t,e) = N^*_a(t,e,\infty)$, and the at-risk processes by
$Y^*_a(t,e,r) = \I\{ T^*_a(e,r) + e \geq t\}$ and
$Y^*_a(t,e) = Y^*_a(t,e,\infty)$, $a=0, 1$ (Fleming and Harrington,
2005).  From the above consistency assumptions, if $t < r$, then
$N^*_a(t,e,r) = N^*_a(t,e)$, $Y^*_a(t,e,r) = Y^*_a(t,e)$, $a=0, 1$.
For $a=0, 1$, let
$\Lambda_a(t, e, r) = \int_0^t \lambda_a(u,e,r) \, du$ be the
cumulative hazard.  Because
$E\{ dN^*_a(t,e,r) | Y^*_a(t, e, r)\} = d\Lambda_a(t, e, r) Y^*_a(t,
e, r)$, $a=0, 1$, it follows that
$\{ dN^*_a(t,e,r) - d\Lambda_a(t, e, r) Y^*_a(t, e, r)\}$, $a = 0, 1$,
are mean-zero counting process increments.  Thus, any linear
combination of these increments over $t, e, r$ can be used to define
unbiased estimating functions in $W^*$ of quantities of interest.  In
Appendix D, we formulate a particular
set of estimating functions such that, given iid potential outcomes
$W^*_i$, $i=1,\ldots,n$, lead to consistent and asymptotically normal
estimators for $\{ \Lambda^b(t), \Lambda^u(t), \theta^T\}^T$,
$\Lambda^k(t) = \int_0^t \lambda^k(u)\, du$, $k=b, u$.  Because
interest focuses on $VE(\tau)$ for $\tau \geq \ell$, estimation of
$\Lambda^u_\ell(t) = \int_0^t \lambda_\ell^u(u)\, du$ and $\zeta( \cdot)$
is not considered.  

For fixed $t$, $0 \leq t \leq L$,
the estimating functions for $\Lambda^b(t)$ and $\Lambda^u(t)$ are,
respectively,
\begin{align}
  &\calE^*_{\Lambda^b}\{ W^*; \Lambda^b(t),\theta\} = \I(t < \calT_C) \left(
  \int_0^{\min(t,\calT_A)} \{ dN^*_0(t,e) - d\Lambda^b(t) Y^*_0(t,
  e)\}  \tilw_0(t,e)\, de \right. \label{eq:Eb} \\*[-0.06in]
&\,\,+ \!\! \left.  \I(t \geq \ell) \int_0^{\min(t-\ell,\calT_A)} \!\!\! \big[ dN^*_1(t,e) - d\Lambda^b(t) \exp\{
  \theta_0+g(t-e-\ell; \theta_1) \I(t-e \geq\ell)\} Y^*_1(t, e) \big]  \tilw_1(t,e)\, 
  de \right), \nonumber
\end{align}
\vspace*{-0.2in}
\begin{equation}
\label{eq:Eu}
\begin{aligned}
  \calE^*_{\Lambda^u}\{ W^*; &\Lambda^u(t),\theta\} = \I(t \geq  \calT_P+\ell) \left(
  \int_0^{\calT_A} \!\!\!\int_{\calT_P}^{\min(t-\ell,\calT_C)}\!\!\big[  dN^*_0(t,e,r ) \right.\\*[-0.12in]
&\hspace*{0.1in}\left.- d\Lambda^u(t) \exp\{  g(t-r-\ell; \theta_1) \I(t -r \geq \ell)\}
Y^*_0(t,e,r) \big]  w_0(t,e,r) \,dr\,  de
\vphantom{\int_0^{\calT_A}  \!\!\!\int_{\calT_P}^{\min(t-\ell,\calT_C)}}\right) \\
&+ \I(t \geq \calT_P) \left( \int_0^{\calT_A}\!\!\!
  \int_{\calT_P}^{\min(t,\calT_C)}\big[ dN^*_1(t,e,r) - d\Lambda^u(t)
  \exp\{ 
  g(t-e-\ell; \theta_1)\} \right.\\*[-0.12in]
&\hspace*{2.1in}\times
\left. \vphantom{\int_0^{\calT_A}\!\!\!\int_{\calT_P}^{\min(t,\calT_C)}}
Y^*_1(t, e,r) \big]  w_1(t,e,r) \I(t \geq r) \,  dr\, de \right),
\end{aligned}
\end{equation}
where $\tilw_a(t, e)$ and $w_a(t, e, r)$, $a=0, 1$, are arbitrary
nonnegative weight functions, specification of which is discussed
later.  The estimating function for $\theta$ is given by
\begin{align}
&\calE^*_\theta\{W^*; \Lambda^b(\cdot), \Lambda^u(\cdot), \theta\}  \nonumber \\*[-0.05in]
 &=\int_\ell^{\calT_C} \!\!\! \int_0^{\min(t-\ell,\calT_A)} \! \dge\!
   \big[ dN^*_1(t,e) - d\Lambda^b(t) \exp\{ \theta_0+g(t-e-\ell; \theta_1) \nonumber \\*[-0.3in]
&\hspace*{3in}\times \I(t-e \geq \ell)\} Y^*_1(t, e) \big]  \tilw_1(t,e)\, de   \nonumber\\*[-0.05in]
 &+\int_{\calT_P+\ell}^L \!
   \int_0^{\calT_A} \!\!\!\int_{\calT_P}^{\min(t-\ell,\calT_C)}\! \dgr\!\big[  dN^*_0(t,e,r ) - d\Lambda^u(t) \exp\{
   g(t-r-\ell; \theta_1) \I(t -r \geq \ell)\} \nonumber \\*[-0.3in]
 &\hspace*{3in}\times Y^*_0(t,e,r) \big]  w_0(t,e,r) \, dr\,  de \label{eq:Et}\\*[-0.05in]
 &+ \int_{\calT_P}^L \!
   \int_0^{\calT_A} \!\!\!\int_{\calT_P}^{\min(t,\calT_C)} \! \dg\! \big[  dN^*_1(t,e,r ) - d\Lambda^u(t) \exp\{
   g(t-e-\ell; \theta_1)\} \nonumber\\*[-0.3in]
 &\hspace*{2.8in}\times
 Y^*_1(t, e,r) \big]  w_1(t,e,r) \I(t \geq r) \,  dr\, de, \nonumber
\end{align}
where $g_\theta(u) = \partial/\partial \theta_1\{g(u;\theta_1)\}$.
Analogous to Yang, Tsiatis, and Blazing (2018),
envisioning (\ref{eq:Eb})-(\ref{eq:Et}) as characterizing a system of
estimating functions
$$\calE^*\{W^*; \Lambda^b(\cdot), \Lambda^u(\cdot),\theta\} =
[\calE^*_{\Lambda^b}\{ W^*; \Lambda^b(t),\theta\},
\calE^*_{\Lambda^u}\{ W^*; \Lambda^u(t),\theta\}, 0 \leq t \leq L,
\calE^*_\theta\{W^*; \Lambda^b(\cdot), \Lambda^u(\cdot),
\theta\}^T]^T,$$ if we could observe $W^*_i$, $i=1\ldots,n$, we would
estimate $d\Lambda^b(\cdot), d\Lambda^u(\cdot), \theta$ by solving the
estimating equations
$\sum_{i=1}^n \calE^*(W_i^*; \Lambda^b(\cdot),
\Lambda^u(\cdot),\theta)\} = 0$.

\vspace*{-0.1in}

\subsection{Identifiability assumptions}
\label{ss:assumptions}

Of course, the potential outcomes $W^*_i$, $i=1,\ldots,n$, are not
observed.  However, we now present assumptions under which we can
exploit  the developments in the last section to derive estimating equations
yielding estimators based on the observed data (\ref{eq:data}).  

Define the indicator
that a participant is observed to be infected at time $t$ by
$dN(t) = \I(U=t,\Delta=1)$, the observed at-risk indicator at $t$ by
$Y(t) = \I(E < t \leq U)$, and
\vspace*{-0.1in}
\begin{align}
I_0(t,e) &= (1-A) \I(E=e) \I(R \geq t), \hspace*{0.2in} I_1(t,e) = A \,
  \I(E=e) \I(R \geq t), \nonumber \\
I_{01}(t,e,r) = & \,(1-A)\I(E=e)\{\I(R=r,\Gamma=1,\Psi=1) +
                  \I(R=r,\Gamma=2,\Psi=1)\}, \label{eq:indicators}\\
&I_{11}(t,e,r) = A\,\I(E=e)\{ \I(R=r,\Gamma=1) + \I(R=r,\Gamma=2)\}. \nonumber
\end{align}
$I_a(t,e)=1$ indicates that a subject entering the trial at time $e$
and randomized to placebo ($a=0$) or vaccine ($a=1$) has not yet been
infected or unblinded by $t$.  For $t>r$, $I_{01}(t,e,r)=1$ indicates that a
subject randomized to placebo at entry time $e$ is unblinded (either
by request or at a PDCV) at time $r$ and crosses over to vaccine at
$r$, and $I_{11}(t,e,r) = 1$ if a subject randomized to vaccine at entry
time $e$ is unblinded at $r$.  Make the consistency assumptions
\begin{align}
I_a(t,e) dN(t) = I_a(t,e) dN_a^*(t,e), \hspace*{0.1in} I_a&(t,e) Y(t) =
  I_a(t,e) Y_a^*(t,e), \hspace*{0.1in} a=0, 1, \nonumber \\
I_{01}(t,e,r)dN(t) = I_{01}(t,e,r)  dN_0^*(t,e,r), &\hspace*{0.1in}
I_{01}(t,e,r)Y(t) = I_{01}(t,e,r)  Y_0^*(t,e,r), \label{eq:consistency}\\
I_{11}(t,e,r) dN(t) = I_{11}(t,e,r) dN_1^*(t,e,r), &\hspace*{0.1in} I_{11}(t,e,r) Y(t) =  I_{11}(t,e,r) Y_1^*(t,e,r).
\nonumber 
\end{align}

We now make assumptions similar in spirit to those adopted in
observational studies.  By randomization, 
\begin{equation}
A \independent (X,E,W^*),
\label{eq:nuc1}
\end{equation}
where we subsume the site indicator $S$ in $X$, and let
$p_A = \pr(A=1)$.  It is realistic to assume that the mix of baseline
covariates changes over the accrual period; e.g., during the trial,
because of lagging accrual of elderly subjects and subjects from
underrepresented groups, an effort was made to increase
participation of these groups in the latter part of the accrual
period.  Accordingly, we allow the distribution of entry time $E$ to
depend on $X$, and denote its conditional density as $f_{E|X}(e | x)$.
We make the no unmeasured confounders assumption \vspace*{-0.1in}
\begin{equation}
E \independent W^* \, | \, X.
\vspace*{-0.1in}
\label{eq:nuc2}
\end{equation}

Define the hazard functions of unblinding in the periods between the Pfizer EUA
and the start of PDCVs and after the start of PDCVs, respectively, as
\begin{equation*}
\begin{aligned}
\lambda_{R,1}(r | X, A, E, W^*) = \lim_{dr \rightarrow 0} \pr(r \leq R <
  r+dr, \Gamma=1 | R \geq r, X, A, E, W^*), \hspace{0.1in} \calT_P \leq r < \calT_U  \\
\lambda_{R,2}(r | X, A, E, W^*) = \lim_{dr \rightarrow 0} \pr(r \leq R <
  r+dr, \Gamma=2 | R \geq r, X, A, E, W^*), \hspace{0.1in} \calT_U \leq r < \calT_C,  \\
\end{aligned}
\end{equation*}
where $\lambda_{R,j}(r | X, A, E, W^*) = 0$ for $r \geq \calT_U$ ($j=1$)
and $r \geq \calT_C$ ($j=2$).  Because the accrual period was short
relative to the length of follow-up, we take these unblinding hazard
functions to not depend on $E$, although including such dependence is
straightforward; and, similar to a noninformative censoring
assumption, to not depend on $W^*$ and write
\vspace*{-0.06in}
\begin{equation}
\lambda_{R,j}(r | X, A, E, W^*) = \lambda_{R,j}(r | X, A),
\hspace*{0.1in} j=1, 2.
\label{eq:nuclam}
\end{equation}
Define
$\calK_R(r | X, A) = \exp[ -\{\Lambda_{R,1}(r | X, A) +
\Lambda_{R,2}(r | X, A)\}]$,
$\Lambda_{R,j}(r | X, A) = \int_{\calT_j}^r \lambda_{R,j}(u | X, A)\,
du$, $\calT_j=\calT_P$ ($j=1$) or $\calT_j=\calT_U$ ($j=2$).
Because $\lambda_{R,1}(r | X, A)$ and
$\lambda_{R,2}(r | X, A)$ are defined on the nonoverlapping intervals
$[\calT_P,\calT_U)$ and $[\calT_U,\calT_C)$, respectively, with
$\calK_{R,j}(r | X, A) = \exp\{-\Lambda_{R,j}(r | X,A)\}$, $j=1,2$, 
\begin{equation*}
\begin{aligned}
\calK_R(r | X, A) &= 1, &r<\calT_P, \\*[-0.06in]
& = \calK_{R,1}(r | X, A),&\calT_P \leq r < \calT_U, \\*[-0.06in]
& = \calK_{R,1}(\calT_U | X,A) \calK_{R,2}(r | X, A),&\calT_U \leq r< \calT_C, \\*[-0.06in]
&= 0, &r \geq \calT_C.
\end{aligned}
\end{equation*}
Finally, define $f_{R,j}(r | X, A) = \calK_R(r |X, A) \lambda_{R,j}(r| X, A)$, $j=1, 2$.  

Let $\pr(\Psi=1 | X, E,\Gamma,R,W^*)$ be the probability that a placebo
participant unblinded at $R$ agrees to receive the Moderna vaccine.
Similar to (\ref{eq:nuclam}), we assume this probability does not
depend on $E, W^*$; moreover, because the unblinding interval
$[\calT_P,\calT_C)$ is very short relative to the length of follow-up,
we assume it does not depend on $R$ but does depend on the unblinding
dynamics at $R$.  Thus, write
\begin{equation}
\pr(\Psi=1 | X, E,\Gamma,R,W^*) =\pr(\Psi=1 | X, \Gamma) =
p_\Psi(X,\Gamma).
\label{eq:nucpsi}
\end{equation}

\vspace*{-0.2in}

\subsection{Observed data estimating equations}
\label{ss:observed}

We now outline, under the assumptions
(\ref{eq:consistency})-(\ref{eq:nucpsi}), which we take to hold
henceforth, how we can develop unbiased estimating equations based on
the observed data yielding consistent and asymptotically normal
estimators for $d\Lambda^b(\cdot), d\Lambda^u(\cdot), \theta$.  The
basic premise is to use inverse probability weighting (IPW) to
probabilistically represent potential outcomes in terms of the
observed data to mimic the estimating functions
(\ref{eq:Eb})-(\ref{eq:Et}).

Considering (\ref{eq:indicators}), define the inverse
probability weights
$$h_0(t,e | X) = (1-p_A) f_{E|X}(e|X) \calK_R(t | X, A=0), \hspace*{0.1in}
h_1(t,e | X) = p_A \,f_{E|X}(e|X) \calK_R(t | X, A=1),$$
\begin{align*}
&\hspace*{0.1in}h_{01}(e, r| X) = (1-p_A) f_{E|X}(e|X) \\*[-0.1in]
&\hspace*{0.4in}\times \{ f_{R,1}(r | X,A=0)  p_\Psi(X,\Gamma=1) + 
f_{R,2}(r | X,A=0)  p_\Psi(X,\Gamma=2)\}, 
\end{align*}
$$h_{11}(e,r|X) = p_A \,f_{E|X}(e|X) \{ f_{R,1}(r | X,A=1)  + f_{R,2}(r | X,A=1) \}.$$
We show in Appendix E that
\begin{align}
&E\left\{\frac{I_0(t,e)dN(t)}{h_0(t,e|X)} \,\middle\vert\,  X,W^*\right\}=dN_0^*(t,e), \hspace*{0.1in}
E\left\{\frac{I_0(t,e)Y(t)}{h_0(t,e|X)} \,\middle\vert\,  X,W^*\right\}=Y_0^*(t,e)   \label{eq:ipw0} \\
&E\left\{\frac{I_1(t,e)dN(t)}{h_1(t,e|X)} \,\middle\vert\,  X,W^*\right\}=dN_1^*(t,e), \hspace*{0.1in}
E\left\{\frac{I_1(t,e)Y(t)}{h_1(t,e|X)} \,\middle\vert\,  X,W^*\right\}=Y_1^*(t,e),   \label{eq:ipw1} \vspace*{-0.1in}
\end{align}
\begin{align}
&E\left\{\frac{I_{01}(t,e,r)dN(t)}{h_{01}(e,r|X)} \,\middle\vert\,  X,W^*\right\}=dN_0^*(t,e,r), \hspace*{0.1in}
E\left\{\frac{I_{01}(t,e,r)Y(t)}{h_{01}(e,r|X)} \,\middle\vert\,  X,W^*\right\}=Y_0^*(t,e,r),   \label{eq:ipw01} \\
&E\left\{\frac{I_{11}(t,e,r)dN(t)}{h_{11}(e,r|X)} \,\middle\vert\,  X,W^*\right\}=dN_1^*(t,e,r), \hspace*{0.1in}
E\left\{\frac{I_{11}(t,e,r)Y(t)}{h_{11}(e,r|X)} \,\middle\vert\,  X,W^*\right\}=Y_1^*(t,e,r).   \label{eq:ipw11} 
\end{align}
To obtain observed data analogs to the estimating functions
(\ref{eq:Eb})-(\ref{eq:Et}), based on the equalities in
(\ref{eq:ipw0})-(\ref{eq:ipw11}), we substitute the IPW expressions in
the conditional expectations on the left hand sides.  Using
(\ref{eq:indicators}) and (\ref{eq:ipw0})-(\ref{eq:ipw1}), the analog to 
(\ref{eq:Eb}) is given by
\begin{equation*}
\begin{aligned}
  &\calE_{\Lambda^b}\{ O ; \Lambda^b(t),\theta\} = \I(t < \calT_C) \left(
  \int_0^{\min(t,\calT_A)} \frac{I_0(t,e)}{h_0(t,e|X)}\{ dN(t) - d\Lambda^b(t) Y(t)\}  \tilw_0(t,e)\, de \right. \\*[-0.06in]
&+\I(t \geq \ell)\int_0^{\min(t-\ell,\calT_A)} \!\!\!\frac{I_1(t,e)}{h_1(t,e|X)}\big[ dN(t) - d\Lambda^b(t) \exp\{
  \theta_0+g(t-e-\ell; \theta_1) \\*[-0.1in]
&\hspace{3.5in}\times \I(t-e \geq \ell)\} Y(t) \big]\left. \tilw_1(t,e)\, de \vphantom{\int_0^{\min(t,\calT_A)} \frac{I_0(t,e)}{h_0(t,e|X)}}\right) 
\end{aligned}
\end{equation*}
\vspace*{-0.15in}
\begin{equation}
\label{eq:Ebobs}
\begin{aligned}
&= \I(t < \calT_C)  \left( \frac{(1-A)\I(R\geq t)}{h_0(t,E|X)}\{ dN(t) - d\Lambda^b(t) Y(t)\}  \tilw_0(t,E)\right. \\*[-0.04in]
&+ \left. \frac{A \I(E+\ell \leq t \leq R) }{h_1(t,E|X)}\big[ dN(t) - d\Lambda^b(t) \exp\{ \theta_0+g(t-E-\ell; \theta_1)\} Y(t) \big]  \tilw_1(t,E) \right).
\end{aligned}
\end{equation}
Likewise, using (\ref{eq:ipw01})-(\ref{eq:ipw11}), the analog to 
(\ref{eq:Eu}) is
\begin{equation*}
\begin{aligned}
  \calE_{\Lambda^u}\{O; &\Lambda^u(t),\theta\} = \I(t \geq \calT_P+\ell) \left(
  \int_0^{\calT_A} \!\!\!\int_{\calT_P}^{\min(t-\ell,\calT_C)} \frac{I_{01}(t,e,r)}{h_{01}(e,r|X)}
\big[  dN(t) \right.\\*[-0.12in]
&\hspace*{1.1in}
\left. \vphantom{\int_0^{\calT_A}\!\!\!\int_{\calT_P}^{\min(t,\calT_C)}}
- d\Lambda^u(t) \exp\{  g(t-r-\ell; \theta_1) \I(t -r\geq \ell)\}  Y(t) \big]  w_0(t,e,r) \,dr\,  de \right)\\
&+ \I(t \geq \calT_P) \left(\int_0^{\calT_A}\!\!\!   \int_{\calT_P}^{\min(t,\calT_C)}
\frac{I_{11}(t,e,r)}{h_{11}(e,r|X)}
\big[ dN(t) - d\Lambda^u(t) \exp\{  g(t-e-\ell; \theta_1)\} \right.\\*[-0.12in]
&\hspace*{1.1in}\times
\left. \vphantom{\int_0^{\calT_A}\!\!\!\int_{\calT_P}^{\min(t,\calT_C)}}
Y(t) \big]  w_1(t,e,r) \I(t \geq r) \,  dr\, de \right)
\end{aligned}
\end{equation*}
\begin{equation}
\label{eq:Euobs}
\begin{aligned}
=\I(t \geq &\calT_P+\ell) \left( \frac{(1-A) \I(t-R \geq \ell) \{\I(\Gamma=1,\Psi=1) + \I(\Gamma=2,\Psi=1)\} }
{h_{01}(E, R|X)} \right. \\*[-0.12in]
&\hspace*{1.1in}\times \left. \vphantom{\frac{(1-A) \I(t \geq R) \I(t-R \geq
    \ell) \{\I(\Gamma=1,\Psi=1) + \I(\Gamma=2,\Psi=1)\} }{h_{01}(E, R|X)}  }
\big[  dN(t) - d\Lambda^u(t) \exp\{  g(t-R-\ell; \theta_1)\} Y(t)
\big]  w_0(t,E,R) \right) \\*[-0.02in]
&+ \I(t \geq\calT_P) \left(\frac{A \I(t > R) \{\I(\Gamma=1) + \I(\Gamma=2)\} }{h_{11}(E, R|X)} \right.\\*[-0.12in]
&\hspace*{1.1in}\left.\vphantom{\frac{A \I(t \geq R) \{\I(\Gamma=1) + \I(\Gamma=2)\} }{h_{11}(E, R|X)}}
\times \big[  dN(t) - d\Lambda^u(t) \exp\{  g(t-E-\ell; \theta_1)\} Y(t) \big]  w_1(t,E,R) \right).
\end{aligned} 
\end{equation}
A entirely similar representation $\calE_\theta\{O; \Lambda^b(\cdot)
\Lambda^u(\cdot), \theta\}$ of  (\ref{eq:Et}) in terms of the observed
data can be deduced and is suppressed for brevity.  

To simplify notation, based on (\ref{eq:Ebobs}), (\ref{eq:Euobs}), and
the analogous expression for (\ref{eq:Et}), define
\begin{align*}
d\tilN^b(t) &= dN(t) \left\{ \frac{(1-A)\I(R\geq t)\tilw_0(t,E) }{h_0(t,E|X)}
  + \frac{A \I(E+\ell \leq t \leq R)\tilw_1(t,E) }{h_1(t,E|X)} \right\} \\
\tilY^b(t) &= Y(t) \left[ \frac{(1-A)\I(R\geq t)\tilw_0(t,E) }{h_0(t,E|X)}
  + \frac{A \I(E+\ell \leq t \leq R)\tilw_1(t,E) }{h_1(t,E|X)}
            \exp\{\theta_0+g(t-E-\ell;\theta_1)\} \right] \\
d\tilN^u(t) &= dN(t) \left[ \frac{(1-A) \I(t - R \geq \ell) \{\I(\Gamma=1,\Psi=1) + \I(\Gamma=2,\Psi=1)\} w_0(t,E,R)}
{h_{01}(E, R|X)} \right.\\
&\hspace{1in}+ \left. \frac{A \I(t > R) \{\I(\Gamma=1) +
  \I(\Gamma=2)\} w_1(t,E,R)}{h_{11}(E, R|X)} \right] \\
\tilY^u(t) &= Y(t) \left[ \frac{(1-A) \I(t - R \geq \ell) \{\I(\Gamma=1,\Psi=1) + \I(\Gamma=2,\Psi=1)\} w_0(t,E,R)}
{h_{01}(E, R|X)} \right.\\*[-0.2in] 
&\hspace{4in}\times \exp\{g(t-R-\ell; \theta_1) \\
&\hspace{1in}+ \left. \frac{A \I(t > R) \{\I(\Gamma=1) +
  \I(\Gamma=2)\} w_1(t,E,R)}{h_{11}(E, R|X)} \exp\{g(t-E-\ell; \theta_1)\right].
\end{align*}
Define also 
\vspace*{-0.1in}
$$Z^b(t) = A \left(\begin{array}{c} 1 \\
                     g_\theta(t-E-\ell)\end{array}\right),\,\,\,
Z^u(t) = A \left(\begin{array}{c} 0 \\
                     g_\theta(t-E-\ell)\end{array}\right) + (1-A) \left(\begin{array}{c} 0 \\
                     g_\theta(t-R-\ell)\end{array} \right).$$
Then it is straightforward that the observed-data estimating functions are
$$\calE_{\Lambda^b}\{ O ; \Lambda^b(t),\theta\} = d\tilN^b(t) -
d\Lambda^b(t) \tilY^b(t), \hspace{0.1in}
 \calE_{\Lambda^u}\{ O ; \Lambda^u(t),\theta\} = d\tilN^u(t) -
d\Lambda^u(t) \tilY^u(t),$$
$$\calE_\theta\{O; \Lambda^b(\cdot) \Lambda^u(\cdot), \theta\}
= \int_0^{\calT_C} Z^b(t)  \{d\tilN^b(t) -d\Lambda^b(t) \tilY^b(t)\}
+ \int_{\calT_P}^L Z^u(t) \{ d\tilN^u(t) - d\Lambda^u(t) \tilY^u(t)\}.$$

Letting $\tilN^b_i(t)$, $\tilN^u_i(t)$, $\tilY^b_i(t)$,
$\tilY^u_i(t)$, $Z^b_i(t)$, and $Z^u_i(t)$ denote evaluation at $O_i$
in (\ref{eq:data}), the foregoing developments lead to the set of
observed-data estimating equations
\begin{align}
&\sumin \{ d\tilN^b_i(t) - d\Lambda^b(t) \tilY^b_i(t) =  0, \hspace*{0.1in}
\sumin \{ d\tilN^u_i(t) - d\Lambda^u(t) \tilY^u_i(t) =  0, \label{eq:esteqLbu} \\*[-0.05in]
\sumin &\left[ \int_0^{\calT_C} Z^b_i(t)  \{d\tilN^b_i(t) -d\Lambda^b(t) \tilY^b_i(t)\}
+ \int_{\calT_P}^L Z^u_i(t) \{ d\tilN^u_i(t) - d\Lambda^u(t) \tilY^u_i(t)\}  \right] = 0. 
\label{eq:esteqnT}
\end{align}
For fixed $\theta$, the estimators for $d\Lambda^b(t)$ and $d\Lambda^u(t)$
are the solutions to the equations in (\ref{eq:esteqLbu}) given by
\vspace*{-0.15in}
\begin{equation}
d\widehat{\Lambda}^b(t) = \left\{ \sumin \tilY^b_i(t) \right\}^{-1} \sumin d\tilN^b_i(t), \hspace*{0.1in}
d\widehat{\Lambda}^u(t) = \left\{ \sumin \tilY^u_i(t) \right\}^{-1}
\sumin d\tilN^u_i(t).
\label{eq:finalestL}
\end{equation}
Substituting these expressions in (\ref{eq:esteqnT}) yields, after some
algebra, the equation
\begin{align}
\sumin \left[ \int_0^{\calT_C} \{  Z^b_i(t)- \overline{Z}^b(t)\} d\tilN^b_i(t) 
+\int_{\calT_P}^L \{  Z^u_i(t)- \overline{Z}^u(t)\}  d\tilN^u_i(t)
  \right] = 0,
\label{eq:finalestT}
\end{align}
$$\overline{Z}^b(t) = 
\left\{ \sumin \tilY^b_i(t) \right\}^{-1} \sumin Z^b_i(t)
\tilY^b_i(t), \hspace*{0.1in}
\overline{Z}^u(t) = 
\left\{ \sumin \tilY^u_i(t) \right\}^{-1} \sumin Z^u_i(t) \tilY^u_i(t).$$

\section{Practical Implementation and Inference}
\label{s:implementation}

Choice of the weight functions $\tilw_0(t,e)$, $\tilw_1(t,e)$,
$w_0(t,e,r)$, and $w_1(t,e,r)$ is arbitrary but can play an important
role in performance of the resulting estimators.  We
recommend taking a fixed value $\widetilde{x}$ of $X$,
e.g., the sample mean, and setting
$\tilw_a(t,e) = h_a(t,e|\widetilde{x})$ and
$w_a(t,e,r) = h_{a1}(e,r | \widetilde{x})$, $a=0,1$, where the latter
does not depend on $t$.  The resulting weights
$h_a(t,e|\widetilde{x})/h_j(t,e|X)$ and
$h_{a1}(e,r | \widetilde{x})/h_{a1}(e,r|X)$, $a=0,1$, are referred to
as stabilized weights (Robins, Hern\'{a}n, and Brumback, 2000), as
  they mitigate the effect of small inverse probability weights that
  can give undue influence to a few observations.  Note that
  dependence of the inverse probability weights on $p_A$ cancels in
  construction of stabilized weights.  Moreover, if there is no
  confounding, in that $\lambda_{R,j}(r|X,A)$, $j=1,2$ in
  (\ref{eq:nuclam}), $f_{E|X}(e | X)$, and $p_\Psi(X,\Gamma)$ do not
  depend on $X$, the stabilized weights are identically equal to one.

  If the ``survival probabilities'' for $R$, $\calK_{R,j}(r|X,A)$, and
  the densities $f_{R,j}(r|X,A)$, $j=1,2$, and $f_{E|X}(e|X)$ in the
  inverse probability weights, which appear in the expressions in the
  estimating equation (\ref{eq:finalestT}), were known,
  (\ref{eq:finalestT}) could be solved to yield an estimator for
  $\theta$ and in particular $\theta_1$ characterizing VE waning.  As
  these quantities are unknown, models must be posited for them,
  leading to estimators that can be substituted in
  (\ref{eq:finalestT}).  We propose the use of Cox proportional
  hazards models for $\lambda_{R,j}(r|X,A)$, $j=1,2$, in
  (\ref{eq:nuclam}), which can be fitted using the data 
$\{X_i, A_i, R_i, \I(\Gamma_i=j)\}$, $i=1\ldots,n$; and for
the hazard of entry time $E$ given $X$, which can be fitted using 
$(E_i, X_i)$, $i=1,\ldots,n$.   A binary, e.g., logistic, regression
model can be used to represent  $p_\Psi(X,\Gamma)$ and fitted using 
$(X_i, \Gamma_i, \Psi_i)$ for $i$ such that $A_i=0$.   

For individual $i$, the stabilized weights involve the quantities
$f_{R,j}(R_i |\widetilde{x}, a)/ f_{R,j}(R_i |X_i, a)$, $j=1,2$,
$a=0, 1$, and $f_{E|X}(E_i | \widetilde{x})/f_{E|X}(E_i | X_i)$.  With
proportional hazards models as above with predictors
$\phi_j(X, \beta_j)$, say, it is straightforward that $f_{E|X}(E_i |
\widetilde{x})/f_{E|X}(E_i | X_i)$ and 
$$f_{R,j}(R_i |\widetilde{x}, a)/ f_{R,j}(R_i |X_i, a) = [ \exp\{
\phi_j(\widetilde{x},\beta_j)\} \calK_R (R_i | \widetilde{x}, a)]/ [
\exp\{ \phi_j(X_i,\beta_j)\} \calK_R (R_i |X_i, a)],$$ where in each
case the baseline hazard cancels from numerator and denominator.
for  Thus, the
estimated stabilized weights involve only the estimated cumulative
hazard functions and estimators for the $\beta_j$, each of which is
root-$n$ consistent and asymptotically normal.

As sketched in Appendix F, with stabilized weights set equal to
one or estimated, the estimating equation (\ref{eq:finalestT}) can be
solved easily via a Newton-Raphson algorithm.  A heuristic argument
demonstrating that $\hattheta$ is asymptotically normal leading to an
expression for its approximate sampling variance using the sandwich
technique is given in Appendix F.

\section{Simulations}
\label{s:sim}

We report on simulation studies demonstrating performance of the
methods, each involving 1000 Monte Carlo replications, based roughly
on the Moderna trial.  We took $p_A = 0.5$ and $\calT_A=12$,
$\calT_P=19$, $\calT_U= 21$, and $\calT_C=31$, where all times are in
weeks, and consider an analysis at calendar time $L=52$ weeks, with
$n=$ 30,000.  In all cases, $g(u,\theta_1) = \theta_1 \I(u > v)$ where
$v = 20$ weeks and $\theta_0 = \log(0.05)$, corresponding to VE = 95\%
prior to time $v$, so that, depending on $\theta_1$, VE potentially
wanes following $v$.  We consider $\theta_1=\log(7)$, corresponding
to VE = 65\% after time $v$, and $\theta_1=0$, corresponding to no
waning.

Because the trial and unblinding process are ongoing, we were not able
to base our generative scenarios on data from the trial.  Owing to the
complexity of the trial and multiple potential sources of confounding,
to facilitate exploration of a range of conditions while controlling
computational complexity and intensity, we focused on several basic
scenarios meant to represent varying degrees of confounding consistent
with our expectations for the most likely sources of such confounding
in the trial.  Specifically, we took $f_{E|X}(e | X)$ and
$\lambda_{R,2}(r|X,A)$ to not depend on $X$ (or $A$ in the latter
case) in any scenario, reflecting mostly random entry and PDCV
unblinding processes.  In scenarios involving confounding, we took
$\lambda_{R,1}(r|X,A)$, corresponding to the period
$[\calT_P,\calT_U)$ in which ``requested unblinding'' occurred, and the
``agreement process'' $p_\Psi(X,\Gamma)$ to depend on $X$, as
described below, reflecting our belief that these processes could be
associated with participant characteristics.

In the first set of simulations, we consider two cases: (i) no
confounding, where all of $\lambda_{R,j}(r|X,A)$, $j=1,2$,
$f_{E|X}(e | X)$, and $p_\Psi(X,\Gamma)$ do not depend on $X$; and
(ii) confounding, where $\lambda_{R,1}(r|X,A)$ and $p_\Psi(X,\Gamma)$
depend on $X$ as above.  In both (i) and (ii), the entry process
$E \sim \calU(0,\calT_A)$, i.e., uniform on $[0,\calT_A]$, and the
unblinding process during PDCVs was $\calU(\calT_U,\calT_C)$; see
below.  In each simulation experiment, for each participant in each Monte
Carlo data set, we first generated $A \sim \text{Bernoulli}(p_{A})$,
two baseline covariates $X_1 \sim \text{Bernoulli}(p_{X_1}=0.5)$ and
$X_2 \sim \N(\mu_{X_2}=45, \sigma^2_{X_2}=10^2)$, and $E$ as above.
To obtain $R$, we generated $G_1$ to be exponential with hazard
$\lambda_{R,1}(r|X,A) = \exp[\tilbeta_{10} + \{\tilbeta_{11} (X_1-p_{X_1})
+\tilbeta_{12}(X_2-\mu_{X_2}) \}(1-A) + \{\tilbeta_{13} (X_1-p_{X_1}) +
\tilbeta_{14}(X_2-\mu_{X_2}) \}A]$, where $\tilbeta_{10} = \log(0.036)$,
corresponding to roughly 7\% unblinding during $[\calT_P,\calT_U)$,
and $(\tilbeta_{11}, \tilbeta_{12},\tilbeta_{13}, \tilbeta_{14}) = (0,0,0,0,0)$
for (i), no confounding, and $(-0.8,-0.08,0.8,0.08)$ for (ii),
confounding.  With $R_1 = \calT_P+G_1$ and
$R_2 \sim \calU(\calT_U,\calT_C)$, we let
$\widetilde{\Gamma} = 1 + \I(R_1 \geq \calT_U)$ and
$\widetilde{R} = R_1 \I(\widetilde{\Gamma}=1) + R_2
\I(\widetilde{\Gamma}=2)$.  We generated $\Psi$ as
Bernoulli$\{p_\Psi(X,\widetilde{\Gamma})\}$,
$p_\Psi(X, \widetilde{\Gamma}) = \mbox{expit}\{\tilgamma_0 + \tilgamma_1 (X_1-p_{X_1}) +
\tilgamma_2 (X_2-\mu_{X_2})+ \tilgamma_3 \widetilde{\Gamma} \}$, $\mbox{expit}(u) = (1+e^{-u})^{-1}$,
where $\tilgamma_0 = 1.4$, corresponding to approximately 80\% agreement
to receive the vaccine by unblinded placebo participants, and
$(\tilgamma_1,\tilgamma_2, \tilgamma_3) =(0,0,-0.1)$ for (i) and $= (-0.8,-0.08,-0.1)$ for (ii).

To generate $U, \Delta$, we first generated $T_0^*(E,R)$ and
$T_1^*(E,R)$ based on (\ref{eq:lam0r})-(\ref{eq:lam1r}), with
$\lambda^b(t) = \lambda^b = \exp\{ \delta_0 + \delta_1 (X_1-p_{X_1})
+\delta_2 (X_2-\mu_{X_2}) + \mathcal{Z} \}$, where
$(\delta_0,\delta_1,\delta_2) = \{ \log(0.0006), 0.4, 0.04 \}$,
leading to approximately a 3\% infection rate for placebo participants
over $L$, and $\mathcal{Z} \sim \N(0, 0.04)$,;
$\lambda^u_\ell(t) = \lambda^u_\ell = \lambda^b$; $\zeta(t) = 0$; and
$\lambda^u(t) = \lambda^u = 1.25 \lambda^b$, so that
$\lambda_a(t,e,r)$ in (\ref{eq:lam0r})-(\ref{eq:lam1r}), $a=0, 1$, are
piecewise constant hazards.  $T_0^*(E,R)$ and $T_1^*(E,R)$ were
obtained via inverse transform sampling.  We then generated $U$
(calendar time) as
$U = E + A T_1^*(E,R) + (1-A)\big[ \I\{T_0^*(E,R) < \widetilde{R}\}
T_0^*(E,R) + \I\{T_0^*(E,R) \geq \widetilde{R} \} \{ \Psi T_0^*(E,R) +
(1-\Psi) T^*_r\}$, where $T^*_r = \widetilde{R} + G_2$ for $G_2$
exponential with hazard $\lambda^b$; infection times for unblinded
placebo participants who decline vaccine are not used in the analysis.
Finally, we set $\Delta = \I(U < L)$, and defined
$R = U \I(U \leq \widetilde{R} ) + \widetilde{R} \I(U >
\widetilde{R})$ and $\Gamma = \widetilde{\Gamma} \I(U > R)$.  Although
we obtained $\Psi$ for all $n$ participants, $\Psi$ is used only when
$A=0$, $\Gamma \geq 1$.

For each combination of (i) and (ii) and (a) $\theta_1 = \log(7)$ and
(b) $\theta_1=0$, we estimated $\theta$ and thus $VE(\tau)$ for
$\tau \leq v$ and $\tau > v$ two ways: taking the stabilized weights
equal to one, so disregarding possible confounding, and with estimated
stabilized weights.  The latter were obtained by fitting proportional
hazards models for entry time $E$ with linear predictor
$\nu_1 X_1 + \nu_2 X_2$ and for $\lambda_{R,j}(r|X,A)$, $j=1,2,$ with
linear predictors
$\beta_{11} X_1 +\beta_{12}X_2 + \beta_{13} A + \beta_{14} X_1 A +
\beta_{15} X_2 A$ and $\beta_{21} X_1 + \beta_{22} X_2$, respectively;
and a logistic regression model for
$p_\Psi(X,\Gamma) = \mbox{expit}\{ (\gamma_{10}+\gamma_{11} X_1 + \gamma_{12} X_2 ) \I
(\Gamma=1) + ( \gamma_{20}+\gamma_{21} X_1 + \gamma_{22} X_2 ) \I
(\Gamma=2) \}$.

\begin{table}[t] 
\caption{Simulation results based on 1000 Monte Carlo
    replications, first scenario. Mean = mean of Monte Carlo estimates, Med =
    median of Monte Carlo estimates, SD = standard deviation of Monte
Carlo estimates, SE = average of
  standard errors obtained via the sandwich technique/delta method,
  Cov = empirical coverage of nominal 95\% Wald confidence
  interval (transformed for $VE$). $VE_{\leq 20} = 1-\exp(\theta_0)$, VE prior to $v=20$
  weeks; $VE_{> 20} = 1-\exp(\theta_0+\theta_1)$, VE after $v=20$
  weeks. True values: (a) $\theta_1 = \log(7) =  1.946$, $VE_{\leq 20}
  = 0.95$, $VE_{>20}=0.65$; (b) $\theta = 0$,  $VE_{\leq 20} = VE_{>20}=0.95$.}
\label{t:results}
\renewcommand{\arraystretch}{0.7} 
  \centering \begin{tabular}{crrcccp{0.02in}rrccc} \hline 
& \multicolumn{5}{c}{Stabilized Weights = 1} &&   \multicolumn{5}{c}{Stabilized Weights Estimated} \\
  & \multicolumn{1}{c}{Mean} & \multicolumn{1}{c}{Med} & SD & SE& Cov &&
\multicolumn{1}{c}{Mean} & \multicolumn{1}{c}{Med} & SD & SE & Cov \\ \hline \\*[-0.05in]
 & \multicolumn{11}{c}{(i), no confounding; (a) $\theta_1=\log(7)$} \\*[0.05in]
$\theta_1$ & 1.961 & 1.935 & 0.310 & 0.308 & 0.955 && 1.983 & 1.959 & 0.303 & 0.310 & 0.957\\
$VE_{\leq 20}$ & 0.950 & 0.953 & 0.019 & 0.019 & 0.952 && 0.950 & 0.952 & 0.019 & 0.019 & 0.953 \\
$VE_{>20}$ & 0.634 & 0.663 & 0.183 & 0.174 & 0.956 && 0.626 & 0.662 & 0.188 & 0.177 & 0.957 \\*[0.1in]

& \multicolumn{11}{c}{(ii), confounding; (a) $\theta_1=\log(7)$} \\*[0.05in]
$\theta_1$ &  2.030 & 2.013 & 0.325 & 0.320 & 0.949 && 1.990 & 1.973 & 0.346 & 0.335 & 0.948 \\
$VE_{\leq 20}$ & 0.951 & 0.953 & 0.019 & 0.018 & 0.958 && 0.951 & 0.952 & 0.019 & 0.019 & 0.955\\
$VE_{>20}$ & 0.614 & 0.647 & 0.199 & 0.185 & 0.948 && 0.619 & 0.665 & 0.201 & 0.186 & 0.941\\*[0.1in]

& \multicolumn{11}{c}{(i), no confounding; (b) $\theta_1=0$} \\*[0.05in]
$\theta_1$ & -0.020 & -0.019 & 0.433 & 0.422 & 0.954 && 0.007 &0.019 & 0.421 & 0.424 & 0.958\\
$VE_{\leq 20}$ & 0.950 & 0.952 & 0.020 & 0.019 & 0.955&& 0.950 & 0.952 & 0.020 & 0.019 & 0.956\\
$VE_{>20}$ & 0.947 & 0.954 & 0.032 & 0.030 & 0.958 && 0.946 & 0.953 & 0.033 & 0.031 & 0.954 \\*[0.1in]

& \multicolumn{11}{c}{(ii), confounding; (b) $\theta_1=0$} \\*[0.05in]
$\theta_1$ & 0.053 & 0.045 & 0.446 & 0.436 & 0.955 && 0.011 & -0.004 & 0.452 & 0.450 & 0.956\\
$VE_{\leq 20}$ & 0.951 & 0.952 & 0.019 & 0.019 & 0.958 && 0.950 & 0.952 & 0.020 & 0.019 & 0.955\\
$VE_{>20}$ & 0.944 & 0.951 & 0.035 & 0.032 & 0.957 && 0.945& 0.954 & 0.036 & 0.033 & 0.952\\*[0.1in] \hline
\end{tabular}
\end{table}

Table~\ref{t:results} presents the results for estimation of
$\theta_1$, dictating waning; $VE_{\leq 20} = 1-\exp(\theta_0)$, VE
prior to $v=20$ weeks; and $VE_{> 20} = 1-\exp(\theta_0+\theta_1)$, VE
after $v=20$ weeks.  Because the Monte Carlo distribution of some of
these quantities exhibited slight skewness, those for the VE
quantities likely due to the exponentiation, we report both Monte
Carlo mean and median.  Estimation of $VE_{\leq 20}$ shows virtually
no bias for both (a) and (b); that for $VE_{> 20}$ in case (a) shows
minimal bias and virtually none for (b).  In all cases, standard
errors obtained via the sandwich technique as outlined in Appendix
F along with the delta method for the VEs track the Monte Carlo
standard deviations.  Under both (i) no confounding and (ii)
confounding, estimation of the stabilized weights appears to have
little consequence for precision of the estimators relative to setting
them to equal to one.  95\% Wald confidence intervals, exponentiated
for the VEs, achieve nominal coverage.  For (b) and each combination
of stabilized weights set equal to one or estimated and (i), no
confounding, and (ii), confounding, we also calculated the empirical
Type I error achieved by a Wald test at level of significance 0.05 for
VE waning addressing the null and alternative hypotheses
$H_0: \theta_1 \leq 0$ versus $H_1: \theta_1 > 0$.  These values are
0.043 and 0.056 when using stabilized weights set equal to one under
(i) and (ii), respectively; the analogous values with estimated
weights are 0.046 and 0.050 under (i) and (ii).

In the first set of simulations, the confounding induced by
our generative choices led to little to no bias in the estimators for
$\theta_1$ and the VEs prior to and after 20 weeks.  Notably, modeling
and fitting of the stabilized weights to adjust for potential
confounding shows little effect relative to setting the stabilized
weights to one.  To the extent that this scenario is a plausible
approximation to actual conditions of the trial, it may be that
confounding will not be a serious challenge for the analysis of VE
waning.  

To examine the ability of the methods with estimated stabilized
weights to adjust for confounding that potentially could be
sufficiently strong to bias results, we carried out additional
simulations under settings (a) $\theta_1=\log(7)$ and (b) $\theta_1=0$
with (ii) confounding in which our choices of generative parameters
induce a stronger association between the potential infection times
and the agreement process.  Specifically, we took instead
$(\delta_0,\delta_1,\delta_2)^T = \{ \log(0.0006), 0.7, 0.07\}^T$ and
$(\tilgamma_0,\tilgamma_1,\tilgamma_2, \tilgamma_3) =(1.4,
-1.0,-0.1,-0.1)$, with all other settings identical to those above.

\begin{table}[t] \caption{Simulation results based on 1000 Monte Carlo
  replications, second scenario.  Entries are as in
  Table~\ref{t:results}.  True values: (a)
  $\theta_1 = \log(7) = 1.946$, $VE_{\leq 20} = 0.95$,
  $VE_{>20}=0.65$; (b) $\theta = 0$,  $VE_{\leq 20} = VE_{>20}=0.95$.}
\label{t:results2}
\renewcommand{\arraystretch}{0.7} 
  \centering \begin{tabular}{crrcccp{0.02in}rrccc} \hline 
& \multicolumn{5}{c}{Stabilized Weights = 1} &&   \multicolumn{5}{c}{Stabilized Weights Estimated} \\
  & \multicolumn{1}{c}{Mean} & \multicolumn{1}{c}{Med} & SD & SE& Cov &&
\multicolumn{1}{c}{Mean} & \multicolumn{1}{c}{Med} & SD & SE & Cov \\ \hline \\*[-0.05in]
 
& \multicolumn{11}{c}{(ii), confounding; (a) $\theta_1=\log(7)$} \\*[0.05in]
$\theta_1$ &  2.125 & 2.100 & 0.315 & 0.299 & 0.925 && 2.009 & 2.008 & 0.346 &  0.325 & 0.942\\
$VE_{\leq 20}$ & 0.952 & 0.953 & 0.017 & 0.016 & 0.970 && 0.950 & 0.952  &0.017 & 0.017 &0.964 \\
$VE_{>20}$ & 0.581 & 0.611 & 0.191 & 0.182 & 0.950 && 0.613 & 0.640 & 0.179 & 0.175 & 0.956\\*[0.1in] 

& \multicolumn{11}{c}{(ii), confounding; (b) $\theta_1=0$} \\*[0.05in]
$\theta_1$ & 0.171 & 0.149 & 0.436 & 0.403 & 0.921 &&  0.050 & 0.053 & 0.447 & 0.426 & 0.955 \\
$VE_{\leq 20}$ & 0.951 & 0.953 & 0.173 & 0.171 & 0.967 && 0.950 & 0.952 & 0.018 & 0.017 & 0.962 \\
$VE_{>20}$ & 0.937 & 0.945 & 0.038 & 0.034 & 0.949 &&  0.942 & 0.949 & 0.034 & 0.032 & 0.950 \\*[0.1in] \hline
\end{tabular}
\end{table}

Table~\ref{t:results2} shows the results.  The estimators for
$\theta_1$ and $VE_{>20}$ are slightly biased when stabilized weights
are set equal to one, although coverage probability for the latter is
at the nominal level.  This feature is mitigated by use of estimated
stabilized weights.  Coverage probability for $\theta_1$ is somewhat
lower than nominal.  Under (b), empirical
Type I error achieved by a Wald test at level of significance 0.05 of 
$H_0: \theta_1 \leq 0$ versus $H_1: \theta_1 > 0$.
is 0.122 when stabilized weights are equal to one,
demonstrating the potential for biased inference; Type I error is
0.065 using estimated stabilized weights, leading to a more reliable
test.  



\section{Discussion}
\label{s:discuss}

We have proposed a conceptual framework based on potential outcomes
for study of VE in which assumptions on biological, behavioral, and
other phenomena are made transparent.  The corresponding statistical
framework combines information from blinded and unblinded
participants over time.  We focus on the setting of the Moderna phase
3 trial, but the principles can be adapted to other settings,
including the blinded crossover design of Follmann et al. (2020).  The
methods provide a mechanism to account for possible confounding.


Through condition (ii) in Section~\ref{s:VE}, (ii)
$E\{\pi_1(t,\tau)|X\}/E\{\pi_0(t)|X\}=q(\tau)$, the methods embed the
assumption that VE is similar across current and emerging viral
variants.  If the analyst is unwilling to adopt an assumption like
condition (ii), then it is not possible to rule out that the data from
the blinded (prior to $\calT_P$) and unblinded (starting at $\calT_P$
phases of the trial reflect very different variant mixtures.  In this
case, calendar time and time since vaccination cannot be disentangled,
and thus it is not possible to evaluate VE solely as a function of
time since vaccination.  However, it may be possible to evaluate the
ratio of infection rates under vaccine at any time $t$ (and thus
variant mixture in force at $t$) after different times since
vaccination $\tau_1$ and $\tau_2$, say, during the unblinded phase of
the trial, namely, $\calI^u_1(t,\tau_1)/\calI^u_1(t,\tau_2)$,
$t \geq \calT_P.$ The infection rates can be estimated based on the
infection status data at time $t$ from vaccinated individuals who
received vaccine at times $t-\tau_1$ and $t-\tau_2$, respectively.
These infection rates and their ratio will reflect information about
the waning of the vaccine itself under the conditions at time $t$, and
in fact this infection rate ratio can be viewed as the ratio of
vaccine efficacies at $\tau_1$ and $\tau_2$.  However, because after
$\calT_C$ information on $\calI^u_0(t)$ will no longer be available,
it is not possible to deduce VE itself for $t \geq \calT_C$.  But if
data external to the trial became available that provide information
on VE at $t$, even for small $\tau$, it may be possible to integrate
this information with that from the infection rates to gain insight
into VE as a function of $\tau$.


\section*{Acknowledgements}

The authors thank Dean Follmann for helpful discussions and insights.  

\section*{References}

 \begin{description}


  \item Baden, L. R., El Sahly, H. M, Essink, B., Kotloff, K.,
  Frey, S., Novak, R., et al. for the COVE Study Group (2020).
  Efficacy and safety of the mRNA-1273 SARS-CoV-2 vaccine.  {\em New
    England Journal of Medicine}, {\tt  https://doi.org/10.1056/NEJMoa2035389}.

\item Fintzi, J. and Follmann, D.  (2021).  Assessing vaccine
  durability in randomized trials following placebo crossover.  
 arXiv preprint arXiv:2101.01295v2.  

\item Fleming, T. R. and Harrington, D. P. (2005).  {\em
    Counting Processes and Survival Analysis}.  New York: Wiley.  

\item Follmann, D., Fintzi, J., Fay, M. P., Janes, H. E., Baden,
  L., El Sahly, H. et al. (2020).  Assessing durability of vaccine
  effect following blinded crossover in COVID-19 vaccine efficacy
  trials.   medRxiv. 2020 Dec 14;2020.12.14.20248137. 
 {\tt https://doi.org/10.1101/2020.12.14.20248137}.

\item Halloran, M. E., Longini, I. M, and Struchiner,
  C. J. (1996).  Estimability and interpretation of vaccine efficacy
  using frailty mixing models. 
{\em American Journal of Epidemiology}, 144, 83--97.

\item Lin, D.-Y., Zeng, D., and Gilbert, P. B.  (2021).
  Evaluating the long-term efficacy of COVID-19 vaccines.
  medRxiv. 2021 Jan 13;2021.01.13.21249779.   \\
{\tt https://doi.org/10.1101/2021.01.13.21249779}. 

\item Longini, I. M. and Halloran, M. E. (1996).  frailty
  mixture model for estimating vaccine efficacy.  {\em Journal of the
    Royal Statistical Society, Series C}, 45, 165--173.

\item Moderna Clinical Study Protocol, Amendment 6, 23 December
  2020, available at \\
  \verb+https://www.modernatx.com/sites/default/files/content_documents/Final%20mRNA-+\\\verb+1273-P301%20Protocol%20Amendment%206%20-%2023Dec2020.pdf+

\item Robins, J. M., Hern\'{a}n, M. A., and Brumback,
  B. (2000).  Marginal structural models and causal inference in
epidemiology.  {\em Epidemiology}, 11, 550--560.

\item Rubin, D. B. (1980).  Bias reduction using
  Mahalanobis-metric matching.  {\em Biometrics}, 36, 293--298.  

\item Yang, S., Tsiatis, A. A., and Blazing, M.  (2018).
  Modeling survival distribution as a function of time to treatment
  discontinuation: A dynamic treatment regime approach.  {\em
    Biometrics}, 74, 900--909. 

\end{description}

\setcounter{equation}{0}
\renewcommand{\theequation}{A.\arabic{equation}}

\section*{Appendix A:  Demonstration of (\ref{eq:Rbt}) and (\ref{eq:tau12})}

We demonstrate that, under the conditions in Section 3 of the main
paper, namely, 
\begin{center}
(i)  $\{\pi_1(t,\tau),\pi_0(t)\}\independent
\{S,c^b(t)\}|X$ and $\{\pi_1(t,\tau),\pi_0(t)\}\independent \{S, c_{1\ell}^u(t),
c^u_1(t)\}|X$,
\end{center} 
\begin{center}
(ii)  $E\{\pi_1(t,\tau)|X\}/E\{\pi_0(t)|X\}=q(\tau)$, 
\end{center}
that (\ref{eq:Rbt}) of the main paper,
\begin{equation}
\R^b(t, \tau) = \frac{\calI_1^b(t,\tau)}{\calI_0^b(t)}= 
\frac{E\{p(t,S)c^b(t)\pi_1(t,\tau)\}}{E\{p(t,S)c^b(t)\pi_0(t)\}}
\label{eq:two}
\end{equation}
does not depend on $t$, and the second equality in (\ref{eq:tau12}) of the main paper, 
\begin{equation}
\frac{\calI_1^u(t,\tau_1)}{\calI_1^u(t,\tau_2)} = 
\frac{E\{p(t,S)c_1^u(t)\pi_1(t,\tau_1)\}}{E\{p(t,S)c_1^u(t)\pi_1(t,\tau_2)\}}=
\frac{\R^b(\tau_1)}{\R^b(\tau_2)}, \hspace*{0.1in} \tau_1, \tau_2 \geq \ell;
\label{eq:three}
\end{equation}
the first equality in (\ref{eq:tau12}) of the main paper follows by an
entirely similar argument.  

We can write (\ref{eq:two}) using condition (i) as
$$\R^b(t, \tau) =\frac{E\big[ E\{p(t,S)c^b(t)|X\} E\{\pi_1(t,\tau)|X\}\big]}
{E\big[ E\{p(t,S)c^b(t)|X\} E\{\pi_0(t)|X\}\big] }.$$
By condition (ii), $E\{\pi_1(t,\tau)|X\} = E\{\pi_0(t)|X\} q(\tau)$;
thus, substituting yields 
$$\R^b(t, \tau) = \frac{E\big[ E\{p(t,S)c^b(t)|X\}
  E\{\pi_0(t)|X\}\big] q(\tau)} {E\big[ E\{p(t,S)c^b(t)|X\} E\{\pi_0(t)|X\}\big]} = q(\tau),$$
so that in fact $q(\tau) = \R^b(\tau)$. 

We can write (\ref{eq:three}) as
$$\frac{E\big[ E\{p(t,S)c_1^u(t)|X\} E\{\pi_1(t,\tau_1)|X\}\big]}
{E\big[ E\{p(t,S)c_1^u(t)|X\} E\{\pi_1(t,\tau_2)|X\}\big]}.$$
Under condition (ii), 
$E\{\pi_1(t,\tau_j)|X\} = E\{\pi_0(t)|X\} q(\tau_j)$, $j=1, 2$; thus,
substituting these equalities yields
$$\frac{E\big[ E\{p(t,S)c_1^u(t)|X\}  E\{\pi_0(t)|X\} q(\tau_1)\big]}
{E\big[ E\{p(t,S)c_1^u(t)|X\} E\{\pi_0(t)|X\} q(\tau_2)\}\big]} =
\frac{q(\tau_1)}{q(\tau_2)} = \frac{ \R^b(\tau_1)} {\R^b(\tau_2)},$$
as required.

\setcounter{equation}{0}
\renewcommand{\theequation}{B.\arabic{equation}}

\section*{Appendix B:  Discussion of Assumptions}

The conceptual framework in Section 3 of the main paper in which we
define vaccine efficacy at a particular time since vaccination relies
on some assumptions.  Of critical importance is the assumption
referred to as condition (ii), namely,
\begin{equation}
E\{\pi_1(t,\tau)|X\}/E\{\pi_0(t)|X\}=q(\tau),
\label{eq:assumption}
\end{equation}
which states that the ratio of transmission probabilities over time within
values of $X$ does not change with time and does not depend on
characteristics in $X$ but depends only on time since vaccination.

As noted in Section 3 of the main paper, in our conceptualization, we
let the individual-specific transmission probabilities $\pi_1(t,\tau)$
and $\pi_0(t)$ depend on $t$ to reflect an evolving mixture of
viral variants as mutations of the virus occur over the course of the
pandemic, under which the overall virulence of virus to which
individuals in the study population may be exposed is changing.  From
this point of view, we can regard time $t$ as a ``proxy'' for this
changing variant mixture and its virulence as the study progresses.
If in fact the overall virulence of the variant mixture does not
change or changes only gradually over time, then it may be reasonable
to take $\pi_1(t,\tau) = \pi_1(\tau)$ and $\pi_0(t) = \pi_0$.  In this
case, the ratio $E\{\pi_1(t,\tau)|X\}/E\{\pi_0(t)|X\}$ in
(\ref{eq:assumption}) is a function only of $\tau$ and $X$.  If
instead the variant mixture does change over the course of the study
in a non-trivial way, taking $\pi_1(t,\tau)$ and $\pi(t)$ not to 
depend on $t$ is untenable.  However, if within the mixture of
variants present at any time $t$ we are willing to assume that the ratio of
transmission probabilities between vaccine and placebo stays in
constant proportion for all variants, it again is reasonable to assume
that $E\{\pi_1(t,\tau)|X\}/E\{\pi_0(t)|X\}$ does not depend on $t$ so
is a function only of $\tau$ and $X$.
 
Under either of these perspectives, for (\ref{eq:assumption}) to hold,
we furthermore must be willing to assume that
$E\{\pi_1(t,\tau)|X\}/E\{\pi_0(t)|X\}$ does not depend on $X$ (in
addition to not depending on $t$) and thus depends only on $\tau$.
Adopting (\ref{eq:assumption}) is similar in spirit to making the assumptions
embodied in many popular models; e.g., a constant odds ratio
over categories in the proportional odds model or a constant hazard
ratio over time in the proportional hazards model.  If
(\ref{eq:assumption}) is violated in that
$E\{\pi_1(t,\tau)|X\}/E\{\pi_0(t)|X\}$ does depend on $X$ (but not on
$t$), the implication for the proposed methods is that, in estimating
VE assuming it depends only on $\tau$, one is estimating
roughly a weighted average of VE as a function of $\tau$ over values
of $X$ in a manner similar to the Mantel-Haenzel method; such an
interpretation is also commonly invoked when the proportional odds or
hazards assumptions do not hold.  

If the analyst is unwilling to adopt an assumption like that in
(\ref{eq:assumption}), then it is not possible to rule out that the
data from the blinded (prior to $\calT_P$) and unblinded (starting at 
$\calT_P$, when unblinding requests commenced following the
Pfizer EUA) phases of the trial reflect very different variant
mixtures.  In this case, calendar time and time since vaccination
cannot be disentangled, and thus it is not possible to evaluate
vaccine efficacy solely as a function of time since vaccination.  In
this setting, however, it may be possible to evaluate the ratio of
infection rates under vaccine at any time $t$ (and thus variant
mixture in force at $t$) after different times since vaccination
$\tau_1 \geq \ell$ and $\tau_2 \geq \ell$, say, during the unblinded phase of the trial,
namely, 
$$\calI^u_1(t,\tau_1)/\calI^u_1(t,\tau_2), \hspace*{0.15in} t \geq \calT_P.$$
The infection rates in this ratio presumably can be estimated based on
the infection status data at time $t$ from vaccinated individuals who
received vaccine at times $t-\tau_1$ and $t-\tau_2$, respectively.
These infection rates and their ratio will reflect information about
the waning of the vaccine itself under the conditions at time $t$, and
in fact this infection rate ratio can be viewed as the ratio of
vaccine efficacies at different values $\tau_1$ and $\tau_2$.
However, because after $\calT_C$ information on $\calI^u_0(t)$ will no longer
be available, it is not possible to deduce vaccine efficacy itself
for $t \geq \calT_C$.  But if data external to the trial became available
that provide information on vaccine efficacy at $t$, even for small
$\tau$ it may be possible to integrate this information with that from
the infection rates to gain insight into vaccine efficacy itself as a
function of $\tau$.



\setcounter{equation}{0}
\renewcommand{\theequation}{C.\arabic{equation}}

\section*{Appendix C: Approximate Equivalence of Hazard Rate and Infection Rate}

As an example, consider $\lambda_0(t,e) = \lambda_0(t,e,\infty)$
defined in (\ref{eq:hazarda}) of the main paper.  From Section 3 of
the main paper, the individual-specific infection rate for an
arbitrary subject in the study population at site $S$ who receives
placebo and is never unblinded ($r=\infty)$ is given by
$p(t,S) c^b(t) \pi_0(t)$.  This quantity is a random variable defined
for the population $\Omega$ with probability
$\{P(\omega): \, \omega \in \Omega\}$, where we view $\omega$ as an
individual in $\Omega$.  Thus, the infection rate for
$\omega \in \Omega$ is
$\iota_0(t)(\omega) = p\{t,S(\omega)\} c^b(t)(\omega)
\pi_0(t)(\omega)$, and the population-level infection rate is given
by
$$E\{\iota_0(t) \} = \int_\Omega \iota_0(t)(\omega)\, dP(\omega).$$
In contrast, the hazard at time $t$ is defined by
$$\lambda_0(t,e) = -\frac{d}{dt} \log\big[ \pr\{T^*_0(e) + e \geq
t\}\big],$$
where 
$$\pr\{T^*_0(e) + e \geq t\} = \int_\Omega \pr\{T^*_0(e)(\omega) + e \geq
t\}\, dP(\omega).$$
If  $\omega$ is at risk of infection at time $t$, then this
individual's hazard of becoming infected at $t$ is given by
$\iota_0(t)(\omega)$. Thus,
$$\pr\{T^*_0(e)(\omega) + e \geq t\} = \exp\left\{ -\int_e^t \,\iota_0(u)(\omega)\, du\right\}.$$
We make the rare infection assumption 
\begin{equation}
\int_0^L \iota_0(t)(\omega)\, du < \epsilon \,\,\, \text{a.s.}
\label{rare}
\end{equation}
Now 
$$\lambda_0(t,e) = \frac{ \int_\Omega G(t)(\omega) \iota_0(t)(\omega)
  \, dP(\omega)}
{\int_\Omega G(t)(\omega)  \, dP(\omega)},$$
where, using the rare infection assumption,
$$G(t)(\omega) = \exp\left\{ -\int_e^t \,\iota_0(u)(\omega)\,
  du\right\}  \geq
\exp(-\epsilon) > 1-\epsilon\,\,\,\, \text{a.s.}$$
Because 
$$\int_\Omega G(t)(\omega) \iota_0(t)(\omega)
  \, dP(\omega) \leq \int_\Omega \iota_0(t)(\omega)   \, dP(\omega)$$
and $G(t)(\omega) > 1-\epsilon$ a.s., 
$$\lambda_0(t,e) \leq \frac{\int_\Omega \iota_0(t)(\omega) \, dP(\omega)}{1-\epsilon}.$$
Moreover, because 
$$\int_\Omega G(t)(\omega) \iota_0(t)(\omega)   \, dP(\omega) >
(1-\epsilon) \int_\Omega \iota_0(t)(\omega)   \, dP(\omega)$$
and $\int_\Omega G(t)(\omega)  \, dP(\omega) \leq 1$, 
$$\lambda_0(t,e)  \geq (1-\epsilon) \int_\Omega \iota_0(t)(\omega)
\, dP(\omega).$$
Thus, 
$$(1-\epsilon) < \frac{\lambda_0(t,e) }{\int_\Omega \iota_0(t)(\omega) \, dP(\omega)}
< (1-\epsilon)^{-1}.$$
Consequently, under the rare infection assumption (\ref{rare}), the
population-level infection rate and the population-level hazard rate are
of the same order of magnitude.  

\setcounter{equation}{0}
\renewcommand{\theequation}{D.\arabic{equation}}

\section*{Appendix D: Derivation of Estimating Functions
(\ref{eq:Eb})-(\ref{eq:Et})}

We present derivations leading to the estimating functions
(\ref{eq:Eb})-(\ref{eq:Et}) based on potential outcomes given in
Section~\ref{ss:potential} of the main paper.  Because interest
focuses on $\tau \geq \ell$, from (\ref{eq:lam0r}) and
(\ref{eq:lam1r}) of the main paper, we are concerned only with
$\Lambda^b(t)$, $\Lambda^u(t)$, and $\theta$. Accordingly, to
determine appropriate linear combinations of the mean-zero counting
process increments
$\{ dN^*_a(t,e,r) - d\Lambda_a(t, e, r) Y^*_a(t, e, r)\}$, $a = 0, 1$,
we must deduce relevant values of $t$, $e$, and $r$, where
$e \leq \calT_A$ and $\calT_P \leq r < \calT_C$ by design.  For $a=0$,
from (\ref{eq:lam0r}) of the main paper, the relevant values are
$t < r$ or $t \geq \ell+r$ and $e \leq \min(t,r)$.  For $a=1$, from
(\ref{eq:lam1r}) of the main paper, $e+\ell \leq t \leq r$ and $t > r$.

Consider for fixed $0 \leq t \leq L$, $a = 0, 1$, integrals of the form
\begin{equation}
\int \!\! \int
\{ dN^*_a(t,e,r) - d\Lambda_a(t, e, r) Y^*_a(t, e, r)\} \, w_a(t, e,
r) \, dr \, de,
\label{six1}
\end{equation}
where $w_a(t,e,r)$ is a non-negative weight function, $a=0, 1$.   We
determine the limits of integration for (\ref{six1}) by considering 
three time periods. 

When $t < \calT_P$, at which point all trial participants are still
blinded, so that $t < r$, (\ref{six1}) for $a=0$ becomes, using
(\ref{eq:lam0r}) of the main paper and the consistency assumptions
below (\ref{eq:lam1r}) of the main paper,
\begin{align}
\int_0^{\min(t, \calT_A)}& \!\! \int_{\calT_P}^{\calT_C} 
\{ dN^*_0(t,e) - d\Lambda^b(t)  Y^*_0(t, e)\} \, w_0(t, e,
r) \, dr \, de \nonumber \\
&= \int_0^{\min(t, \calT_A)} 
\{ dN^*_0(t,e) - d\Lambda^b(t)  Y^*_0(t, e)\} \,\tilw_0(t, e) \, de,
\label{seven1}
\end{align}
where for $t <\calT_P$ 
$$\tilw_0(t,e) = \int_{\calT_P}^{\calT_C} w_0(t,e,r)\, dr.$$
For $a=1$, $\ell \leq t < \calT_P$ shows that (\ref{six1})
becomes, using  (\ref{eq:lam1r}) of the main paper, 
\begin{align}
\int_0^{\min(t-\ell, \calT_A)}& \!\! \int_{\calT_P}^{\calT_C} 
\big[ dN^*_1(t,e) - d\Lambda^b(t)  \exp\{\theta_0 + g(t-e-\ell; \theta_1)\}
Y^*_1(t, e)\big] \,w_1(t, e,r) \,dr \, de \nonumber \\
&= \int_0^{\min(t-\ell, \calT_A)} \big[ dN^*_1(t,e) - d\Lambda^b(t)
  \exp\{\theta_0 + g(t-e-\ell; \theta_1)\} Y^*_1(t, e)\big]\,   \tilw_1(t,e)\, de, 
\label{seven2}
\end{align}
where for $t < \calT_P$ 
$$\tilw_1(t,e) = \int_{\calT_P}^{\calT_C} w_1(t,e,r)\, dr.$$

Next consider $\calT_P \leq t < \calT_C$; at times in this interval,
some participants are still blinded while others have become
unblinded.  We consider both $t < r$, so before unblinding, and $t
\geq r$, after unblinding at time $r$.   First consider (\ref{six1})
with $a=0$.  For $t < r$, (\ref{six1}) becomes
\begin{align}
\int_0^{\calT_A}& \!\! \int_{t}^{\calT_C} 
\{ dN^*_0(t,e) - d\Lambda^b(t)  Y^*_0(t, e)\} \, w_0(t, e,
r) \, dr \, de \nonumber \\
&= \int_0^{\calT_A} 
\{ dN^*_0(t,e) - d\Lambda^b(t)  Y^*_0(t, e)\} \,\tilw_0(t, e) \, de,
\label{seven3}
\end{align}
where for $\calT_P \leq t < \calT_C$
$$\tilw_0(t,e) = \int_t^{\calT_C} w_0(t,e,r)\, dr.$$
Similarly, for $a=1$, $t < r$, (\ref{six1}) becomes
\begin{equation}
\int_0^{\calT_A} 
\big[ dN^*_1(t,e) - d\Lambda^b(t)  \exp\{\theta_0 + g(t-e-\ell; \theta_1)\} Y^*_1(t, e)\big] \,\tilw_1(t, e) \, de,
\label{seven5}
\end{equation}
where for $\calT_P \leq t < \calT_C$
$$\tilw_1(t,e) = \int_t^{\calT_C} w_1(t,e,r)\, dr.$$
Continuing to consider $\calT_P +\ell \leq t < \calT_C$, now take $t \geq
r$.  For $a=0$, (\ref{six1}) becomes
\begin{align}
\int_0^{\calT_A} \!\! \int_{\calT_P}^{t-\ell} 
\big[ dN^*_0(t,e,r) - d\Lambda^u(t) \exp\{g(t-r-\ell; \theta_1) \I(t -r\geq \ell)\}  Y^*_0(t, e, r)\big] 
\, w_0(t, e,  r) \, dr \, de,
\label{seven4}
\end{align}
For $a=1$,  (\ref{six1}) becomes
\begin{align}
\int_0^{\calT_A} \!\! \int_{\calT_P}^{t} 
\big[ dN^*_1(t,e,r) - d\Lambda^u(t) \exp\{g(t-e-\ell; \theta_1)\}  Y^*_1(t, e, r)\big] 
\, w_1(t, e,  r) \I(t \geq r) \, dr \, de,
\label{seven6}
\end{align}

Finally, consider $t \geq \calT_C$; these are times where all
participants are unblinded. Thus, when $a=0$, (\ref{six1}) equals
\begin{align}
\int_0^{\calT_A} \!\! \int_{\calT_P}^{\min((t-\ell,\calT_C)} 
\big[ dN^*_0(t,e,r) - d\Lambda^u(t) \exp\{g(t-r-\ell; \theta_1) \I(t -r \geq \ell)\}  Y^*_0(t, e, r)\big] 
\, w_0(t, e,  r) \, dr \, de,
\label{seven7}
\end{align}
and when $a=1$ equals
\begin{align}
\int_0^{\calT_A} \!\! \int_{\calT_P}^{\calT_C} 
\big[ dN^*_1(t,e,r) - d\Lambda^u(t) \exp\{g(t-e-\ell; \theta_1)\}  Y^*_1(t, e, r)\big] 
\, w_1(t, e,  r) \I(t \geq r) \, dr \, de.
\label{seven8}
\end{align}

Combining (\ref{seven1})-(\ref{seven5}) yields estimating function
$\calE_{\Lambda^b}\{ W^*; \Lambda^b(t),\theta\}$ in (\ref{eq:Eb}) of
the main paper.  Combining (\ref{seven4})-(\ref{seven8}) yields 
$\calE_{\Lambda^u}\{ W^*; \Lambda^u(t),\theta\}$ in (\ref{eq:Eu}) of
the main paper.   Estimating function $\calE_\theta\{W^*;
\Lambda^b(\cdot) \Lambda^u(\cdot), \theta\}$ arises through similar
considerations, integrating over $t$ and differentiating with respect
to $\theta_0$ and $\theta_1$.  

\setcounter{equation}{0}
\renewcommand{\theequation}{E.\arabic{equation}}

\section*{Appendix E: Demonstration of (\ref{eq:ipw0})-(\ref{eq:ipw11}) }

We make the assumptions (\ref{eq:consistency})-(\ref{eq:nucpsi}) in
Section~\ref{ss:assumptions} of the main paper.  Here, we show the
first equalities in (\ref{eq:ipw0}) and (\ref{eq:ipw01}), i.e.,
\begin{equation}
E\left\{\frac{I_0(t,e)dN(t)}{h_0(t,e|X)} \,\middle\vert\,  X,W^*\right\}=dN_0^*(t,e)
\label{ipw0}
\end{equation}
and 
\begin{equation}
E\left\{\frac{I_{01}(t,e,r)dN(t)}{h_{01}(e,r|X)} \,\middle\vert\,  X,W^*\right\}=dN_0^*(t,e,r).
\label{ipw01}
\end{equation}
Demonstration of the other equalities in
(\ref{eq:ipw0})-(\ref{eq:ipw11}) follows by analogous arguments.  

We first show (\ref{ipw0}).  By the consistency assumption
(\ref{eq:consistency}) in the main paper, the left hand side of
(\ref{ipw0}) is equal to 
$$E\left\{\frac{I_0(t,e)dN_0^*(t,e)}{h_0(t,e|X)} \,\middle\vert\,
  X,W^*\right\}
= \frac{dN^*_0(t,e)}{h_0(t,e|X)} E\{ I_0(t,e) | X, dN^*_0(t,e)=1,
W^*\}.$$
The result follows if we show that
\begin{equation}
E\{ I_0(t,e) | X, dN^*_0(t,e)=1, W^*\} = h_0(t,e|X).
\label{two1}
\end{equation}
By (\ref{eq:indicators}) of the main paper, the left hand side of
(\ref{two1}) is computed as
\begin{align}
\pr\{E=e&| X, dN^*_0(t,e)=1, W^*\} \label{two3}\\*[-0.06in]
&\times \pr\{ A=0 | X,  E, dN^*_0(t,e)=1, W^*\} \label{two2}\\*[-0.06in]
&\times \pr\{ R > t | X, A=0,  dN^*_0(t,e)=1, W^*\}, \label{two4}
\end{align}
where we have used the assumption discussed above (\ref{eq:nuclam}) in
the main paper in (\ref{two4}).   By (\ref{eq:nuc1}) of the main
paper, (\ref{two2}) is equal to $\pr(A=0) = 1- p_A$.  By
(\ref{eq:nuc1}) and (\ref{eq:nuc2}) of the main paper, (\ref{two3}) is
equal to $f_{E|X}(e|X)$.  The proof will be complete by showing that
(\ref{two4}) is equal to $\calK_R(t|X, A=0)$.  

To demonstrate this, we consider $t < \calT_P$,
$\calT_p \leq t < \calT_U$, $\calT_U \leq t < \calT_C$, and
$t \geq \calT_C$ in turn.  Clearly (\ref{two4}) is equal to 1 for
$t < \calT_P$.  Because the estimating function using $I_0(t,e)$ is
defined only for $t < \calT_C$, we need not consider $t \geq \calT_C$.
Thus, we need only consider the cases $\calT_p \leq t < \calT_U$ and
$\calT_U \leq t < \calT_C$.  For $\calT_p \leq t < \calT_U$, we write
(\ref{two4}) as a product integral as in Anderson et al. (1993) and
use an argument similar to that in (8.72)-(8.77) of Tsiatis et
al. (2020):
\begin{align}
&\prod_{\calT_P \leq w < t} [1 - \pr \{ w \leq R <w+dw | R \geq w, X,
A=0, dN^*_0(t,e)=1, W^*\}] \nonumber\\
&= \prod_{\calT_P \leq w < t} [1 - \pr \{ w \leq R <w+dw, \Gamma=1 | R \geq w, X,
A=0, dN^*_0(t,e)=1, W^*\}] \label{three1} \\
&=\prod_{\calT_P \leq w < t} [1 - \lambda_{R,1}\{w| X,A=0,  dN^*_0(t,e)=1, W^*\} dw] \nonumber \\
&=\prod_{\calT_P \leq w < t} \{1 - \lambda_{R,1}(w| X,A=0) \, dw\} \label{three2}\\
&=\exp\left\{ -\int_{\calT_P}^t \lambda_{R,1}(w| X,A=0) \, dw\right\} =
  \calK_{R,1}(t |  X,A=0) = \calK_{R}(t |  X,A=0), \nonumber
\end{align}
where (\ref{three1}) follows because, if $ dN^*_0(t,e)=1$ and $A=0$,
then the individual could not have been infected before time $t$, and
thus for $\calT_P \leq t < \calT_U$, the only way $R$ could fall
between $w$ and $w+dw$ is if s/he were unblinded during this period,
in which case $\Gamma=1$.  (\ref{three2}) holds because of assumption
(\ref{eq:nuclam}) of the main paper.  Thus, (\ref{two4}) holds for
$\calT_P \leq t < \calT_U$.  Finally, for $\calT_U \leq t < \calT_C$,
write (\ref{two4}) as 
\begin{align}
\pr\{ R &\geq \calT_U | X, A=0, dN^*_0(t,e)=1, W^*\} \label{four1} \\
&\times \pr\{ R > t |  R \geq \calT_U, X, A=0, dN^*_0(t,e)=1, W^*\}.  \label{four2}
\end{align}
From the previous argument, (\ref{four1}) is equal to
$\calK_{R,1}(\calT_U|X, A=0)$, and (\ref{four2}) can be written as a 
product integral, namely, 
$$\prod_{\calT_U \leq w < t} [1 - \pr \{ w \leq R <w+dw | R \geq w, X,
A=0, dN^*_0(t,e)=1, W^*\}],$$
where, using an argument analogous to that above, (\ref{four2}) can be
shown to be equal to $\calK_{R,2}(t|X, A=0)$.  Thus the product of
(\ref{four1}) and (\ref{four2}) is equal to 
$\calK_{R,1}(\calT_U|X, A=0) \calK_{R,2}(t|X, A=0) = \calK_R(t | X,
A=0)$ for $\calT_U \leq t < \calT_C$, completing the proof.  

We now show (\ref{ipw01}).   By the consistency assumption
(\ref{eq:consistency}) in the main paper, the left hand side of
(\ref{ipw01}) is 
$$E\left\{\frac{I_{01}(t,e,r)dN_0^*(t,e,r)}{h_{01}(e,r|X)} \,\middle\vert\,
  X,W^*\right\}
= \frac{dN^*_0(t,e,r)}{h_{01}(e,r|X)} E\{ I_{01}(t,e,r) | X, dN^*_0(t,e,r)=1,
W^*\}.$$
The result will follow if we can show that 
\begin{equation}
E\{ I_{01}(t,e,r) | X, dN^*_0(t,e,r)=1, W^*\} = h_{01}(e,r|X).
\label{five1}
\end{equation}
By (\ref{eq:indicators}) of the main paper, the left hand side of
(\ref{five1}) is computed as
\begin{align}
\pr\{&E=e| X,  dN^*_0(t,e,r)=1, W^*\} \label{five3}\\*[-0.06in]
&\times \pr\{ A=0  | X, E, dN^*_0(t,e,r)=1, W^*\} \label{five2} \\*[-0.06in]
&\times \Big[ \pr\{ R=r,\Gamma=1|X, A=0,  dN^*_0(t,e,r)=1, W^*\} \label{five4}\\*[-0.06in]
&\hspace{0.2in}\times\pr\{\Psi=1|X, A=0,  \Gamma=1,dN^*_0(t,e,r)=1,  W^*\} \label{five5}\\*[-0.06in]
&+ \pr\{ R=r,\Gamma=2|X, A=0,  dN^*_0(t,e,r)=1, W^*\} \label{five6}\\*[-0.06in]
&\hspace{0.2in}\times\pr\{\Psi=1|X, A=0,  \Gamma=2,dN^*_0(t,e,r)=1,  W^*\} \Big].\label{five7}
\end{align}
As in the proof of (\ref{ipw0}), (\ref{five2}) is equal to $(1-p_A)$,
and (\ref{five3}) is equal to $f_{E|X}(e|X)$.  By definition, $R$ is
only defined for values of $r$ between $\calT_P$ and $\calT_C$.  For
$\calT_P \leq r < \calT_U$, $\Gamma$ must be equal to 1, in which case
the product of (\ref{five6}) and (\ref{five7}) is equal to zero.  For
$\calT_U \leq r < \calT_C$, $\Gamma$ must equal to 2, in which case
the product of (\ref{five4}) and (\ref{five5}) is equal to zero.   By
assumption (\ref{eq:nucpsi}) of the main paper, (\ref{five5}) is equal
to $p_\Psi(X, A=0, \Gamma=1)$, whereas (\ref{five4}) can be written as a
product integral
\begin{align}
&\prod_{\calT_P \leq w < r} [1 - \pr \{ w \leq R <w+dw | R \geq w, X,
A=0, dN^*_0(t,e,r)=1, W^*\}] \nonumber\\*[-0.1in]
&\hspace{1in}\times \pr\{r \leq R \leq r+dr, \Gamma=1 | R\geq r, X,
A=0, dN^*_0(t,e,r)=1, W^*\} \nonumber\\
&=\prod_{\calT_P \leq w < r} [1 - \pr \{ w \leq R <w+dw,\Gamma=1 | R \geq w, X,
A=0, dN^*_0(t,e,r)=1, W^*\}] \nonumber\\*[-0.1in]
&\hspace{1in}\times \pr\{r \leq R \leq r+dr, \Gamma=1 | R\geq r, X,
A=0, dN^*_0(t,e,r)=1, W^*\} \label{six11} \\
&=\prod_{\calT_P \leq w < r} [1 - \lambda_{R,1}\{w| X,A=0, dN^*_0(t,e,r)=1, W^*\} dw] \nonumber \\*[-0.1in]
&\hspace*{1in}\times \lambda_{R,1}\{r| X,A=0, dN^*_0(t,e,r)=1, W^*\}  dr \nonumber \\
&=\prod_{\calT_P \leq w < r} \{1 - \lambda_{R,1}(w| X,A=0)\, dw\} \lambda_{R,1}(r| X,A=0) \, dr
\label{six2}\\
&=\exp\left\{ -\int_{\calT_P}^r \lambda_{R,1}(w| X,A=0) \, dw\right\} \lambda_{R,1}(r| X,A=0) \, dr
= f_{R,1}(r|X,A=0), \nonumber
\end{align}
where (\ref{six11}) follows because, if $dN^*_0(t,e,r)=1$ and $A=0$, then
the individual could not have been infected before time $r$.  This
implies that the only way $R$ could fall between $w$ and $w+dw$, for
$w<r$, is if unblinding occurred in this period, in which case
$\Gamma=1$.  (\ref{six2}) holds because of assumption
(\ref{eq:nuclam}) of the main paper.  Thus, (\ref{five1}) holds for
$\calT_P \leq r < \calT_U$.  Analogous arguments can be used to show
that, when $\calT_U \leq t < \calT_C$, the product of (\ref{five6})
and (\ref{five7}) is equal to $f_{R,2}(r|X, A=0) p_\Psi(X, A=0,
\Gamma=2)$, thus demonstrating that (\ref{five1}) holds for $\calT_U
\leq r < \calT_C$, completing the proof.  

\setcounter{equation}{0}
\renewcommand{\theequation}{F.\arabic{equation}}

\section*{Appendix F:  Implementation and Large Sample Properties}


We present a heuristic argument to establish the large-sample
properties of the estimator $\hattheta$ solving  (\ref{eq:finalestT})
of the main paper, namely, 
\begin{align}
\sumin \left[ \int_0^{\calT_C} \{  Z^b_i(t)- \overline{Z}^b(t)\} d\tilN^b_i(t) 
+\int_{\calT_P}^L \{  Z^u_i(t)- \overline{Z}^u(t)\}  d\tilN^u_i(t)
  \right] = 0,
\label{esteqn}
\end{align}
$$\overline{Z}^b(t) = 
\left\{ \sumin \tilY^b_i(t) \right\}^{-1} \sumin Z^b_i(t)
\tilY^b_i(t), \hspace*{0.1in}
\overline{Z}^u(t) = 
\left\{ \sumin \tilY^u_i(t) \right\}^{-1} \sumin Z^u_i(t) \tilY^u_i(t).$$
The estimating equation (\ref{esteqn}) can be written equivalently as
\begin{equation}
\label{esteqnalt}
\begin{aligned}
\sumin &\left[ \int_0^{\calT_C} \{  Z^b_i(t)- \overline{Z}^b(t)\}   \{d\tilN^b_i(t) - d\Lambda^b(t) \tilY^b_i(t) \} \right.\\
&\hspace*{0.3in}\left.+\int_{\calT_P}^L \{  Z^u_i(t)- \overline{Z}^u(t)\} \{ d\tilN^u_i(t) - d\Lambda^u(t) \tilY^u_i(t) \}
  \right] = 0,
\end{aligned}
\end{equation}
which follows because 
$$\sumin \{  Z^k_i(t)- \overline{Z}^k(t)\}  \tilY^k_i(t) = 0,
\hspace*{0.1in} k=b, u.$$
Letting $\mu^k(t)$ be the limit in probability of $\overline{Z}^k(t)$
$k=b, u$, then the left hand side of (\ref{esteqnalt}) can be written
as
\begin{align}
&\sumin \left[ \int_0^{\calT_C} \{  Z^b_i(t)- \mu^b(t)\}
  \{d\tilN^b_i(t) - d\Lambda^b(t) \tilY^b_i(t) \} \right. \nonumber \\
&\hspace*{0.6in}\left.+\int_{\calT_P}^L \{  Z^u_i(t)- \mu^u(t)\} \{ d\tilN^u_i(t) - d\Lambda^u(t) \tilY^u_i(t) \}
  \right] \label{esteqnmu}\\
-&\sumin
\left[ \int_0^{\calT_C} \{ \overline{Z}^b(t)-\mu^b(t)\}
  \{d\tilN^b_i(t) - d\Lambda^b(t) \tilY^b_i(t) \} \right. \nonumber\\
&\hspace*{0.6in}\left.+\int_{\calT_P}^L \{
  \overline{Z}^u(t)-\mu^u(t)\} \{ d\tilN^u_i(t) - d\Lambda^u(t)
  \tilY^u_i(t) \} \label{ignore}
  \right] = 0.
\end{align}
Because $E \{d\tilN^k_i(t) - d\Lambda^k(t) \tilY^k_i(t) \} =0$, and
$\{ \overline{Z}^k(t)-\mu^k(t)\}$ converges in probability to zero
$k = b, u$, (\ref{ignore}) is a small order term that can be ignored
in the sense that $n^{-1/2} \times (\ref{ignore})$ converges in
probability to zero. Thus, solving (\ref{esteqnalt}) is asymptotically
equivalent to setting (\ref{esteqnmu}) equal to zero.  Letting
$\theta^{(0)}$ denote the true value of $\theta$ under the assumption that
the semiparametric model (\ref{eq:semimodel}) of the main paper is
correctly specified, then (\ref{esteqnmu}) is a sum of mean-zero
independent and identically distributed (iid) terms
\begin{equation}
\label{psii}
\begin{aligned}
\psi(O_i; \theta) = \int_0^{\calT_C} & \{  Z^b_i(t)- \mu^b(t)\}
  \{d\tilN^b_i(t) - d\Lambda^b(t) \tilY^b_i(t) \}  \\
&+ \int_{\calT_P}^L \{  Z^u_i(t)- \mu^u(t)\} \{ d\tilN^u_i(t) - d\Lambda^u(t) \tilY^u_i(t) \},
\end{aligned}
\end{equation}
where $E\{ \psi(O_i; \theta^{(0)})\}=0$.   Thus, the estimator
$\hattheta$ solving the asymptotically equivalent estimating equation
$$\sumin \psi(O_i; \theta) = 0$$
satisfies, by a standard Taylor series expansion, 
$$0=\sum_{i=1}^n \psi(O_i,\hattheta)\approx \sum_{i=1}^n
  \psi(O_i,\theta^{(0)})+
\left\{\sum_{i=1}^n\frac{\partial \psi(O_i,\theta_0)}{\partial
    \theta^T}\right\}(\hattheta-\theta^{(0)}).$$
As a consequence,
\begin{equation}
n^{1/2}(\hattheta-\theta^{(0)})=\left[-E\left\{\frac{\partial \psi(O_i,\theta^{(0)})}{\partial
    \theta^T}\right\}\right]^{-1}n^{-1/2}\sum_{i=1}^n\psi(O_i,\theta^{(0)}) +o_P(1), 
\label{convdist}
\end{equation}
which implies that $\hattheta$ is  asymptotically normal with mean zero
and covariance matrix
\begin{equation}\label{avar}
\left[-E\left\{\frac{\partial \psi(O_i,\theta^{(0)})}{\partial
      \theta^T}\right\}\right]^{-1}\mbox{var}\{\psi(O_i,\theta^{(0)})\}\left(\left[-E\left\{\frac{\partial
      \psi(O_i,\theta^{(0)})}{\partial \theta^T}\right\}\right]^{-1}\right)^T,
\end{equation}
where $\mbox{var}\{\psi(O_i,\theta^{(0)})\}=E
\{\psi(O_i,\theta^{(0)}) \psi(O_i,\theta^{(0)})^T\}$.

An estimator for the asymptotic variance (\ref{avar})  can be obtained
as follows.   The term $\mbox{var}\{\psi(O_i,\theta^{(0)})\}$ can be
  estimated by 
$$\widehat{\mbox{var}}\{\psi(O_i,\theta^{(0)})\}=n^{-1}\sum_{i=1}^n
\widehat{\psi}_i(\hattheta)\widehat{\psi}_i(\hattheta)^T,$$ where
$\widehat{\psi}_i(\hattheta)$ is an estimator for $\psi(O_i,\theta^{(0)})$
obtained by substituting (i) $\overline{Z}^k(t)$ for $\mu^k(t)$,
$k = b , u$; (ii) $d\widehat{\Lambda}^k(t)$ in (\ref{eq:finalestL}) of
the main paper for $d\Lambda^k(t)$, $k = b, u$; and (iii) $\hattheta$ for
$\theta^{(0)}$.  An estimator for $$E\left\{\frac{\partial
    \psi(O_i,\theta^{(0)})}{\partial \theta^T}\right\}$$
is obtained by substitutions (i)--(iii) in this expression and
averaging over $i$, leading to 
\begin{equation}\label{gradest}
\widehat{E}\left\{\frac{\partial  \psi(O_i,\theta^{(0)})}{\partial \theta^T}\right\}=
-n^{-1}\sum_{i=1}^n \left\{ \int_0^{\calT_C} V^b(t)\,  d\tilN^b_i(t)
  + \int_{\calT_P}^L V^u(t)\,  d\tilN^u_i(t) \right\},
\end{equation}
where
\begin{align*}  
V^k(t) = \frac{ \sumin \{  Z^k_i(t)- \overline{Z}^k(t)\}   \{  Z^k_i(t)- \overline{Z}^k(t)\} 
^T \tilY^k_i(t)}{ \sumin \tilY^k_i(t)}, \hspace*{0.15in} k=b, u.
\end{align*}
The resulting sandwich estimator for the large sample covariance
matrix of
$\hattheta$  is then given by
\begin{equation}\label{varest}
  \left[\widehat{E}\left\{\frac{\partial
      \psi(O_i,\theta^{(0)})}{\partial \theta^T}\right\}\right]^{-1}
  \widehat{\mbox{var}}\{\psi(O_i,\theta^{(0)})\}
  \left[\widehat{E}\left\{\frac{\partial
      \psi(O_i,\theta^{(0)})}{\partial \theta^T}\right\}\right]^{-1}.
\end{equation}

The foregoing developments take the inverse probability weights and
thus the stabilized weights to be known.  If models for
$\lambda_{R,j}(r|X,A)$, $j=1,2$, $f_{E|X(}(e|X)$, and
$p_\Psi(X,\Gamma)$ are posited and fitted and substituted in
(\ref{esteqn}), then the large sample distribution of
$n^{1/2}(\hattheta-\theta^{(0)})$ would be considerably more
complicated.  In simulations, we have observed that standard errors
and confidence intervals based on (\ref{convdist}) and (\ref{varest})
reflect the true sampling variation in that their numerical values are
consistent with the Monte Carlo sampling variation and confidence
intervals achieve the nominal level of coverage. 
An alternative strategy to obtaining approximate
standard errors and confidence intervals would be to use a
nonparametric bootstrap.

The result (\ref{convdist})  suggests a Newton-Raphson iterative
scheme for solving the estimating equation (\ref{esteqn}).  
Letting $\theta_{(0)}$ be an initial value for $\theta$ and
$\theta_{(m)}$ be the value at the  $m$th
iteration, compute the update by 
$$\theta_{(m+1)}=\theta_{(m)} -
\left[\widehat{E}\left\{\frac{\partial  \psi(O_i,\theta_{(m)})}{\partial \theta^T}\right\}\right]^{-1}
\widehat{\psi}_i(\theta_{(m)}).$$ 
This scheme is iterated until some convergence criterion is satisfied.

\section*{Appendix References}

\begin{description}

\item Anderson, P. K., Borgan, \O., Gill, R. D., and Keiding,
  N. (2993).  {\em Statistical Methods Based on Counting
    Processes}. New York: Springer. 

\item Tsiatis, A. A., Davidian, M., Holloway, S. T., and Laber,
  E. B. (2020).  {\em Dynamic Treatment Regimes: Statistical Methods
    for Precision Medicine}.  Boca Raton, FL: Chapman and Hall/CRC
  Press.

\end{description}

\end{document}